\documentclass[12pt,letterpaper]{article}
\usepackage[usenames, dvipsnames]{xcolor}
\usepackage{jheppub}

\usepackage{tocloft}
\usepackage[]{todonotes}
\usepackage{bbold}
\usepackage{graphicx}
\usepackage{verbatim}
\usepackage[mathscr]{euscript}
\usepackage{slashed}
\usepackage{graphicx}
\usepackage{mathdots}
\usepackage{caption}

\usepackage[labelformat=simple]{subcaption}

\usepackage{enumerate}
\usepackage{bbm}
\usepackage{psfrag}
\usepackage{subfiles}
\usepackage{psfrag}
\usepackage{relsize}
\usepackage{pgfplots}
\usepackage{tikzscale}

\usepackage{bbm} 					
\usepackage{slashed} 				
\usepackage{graphicx}				
\usepackage{subcaption}			
\usepackage{psfrag}				
\usepackage{tensor}				
\usepackage{fouridx}				
\usepackage{bm}					
\usepackage{mdframed}				
\usepackage{multirow}				
\usepackage{soul}					
\usepackage{bbold}				
\usepackage{multicol}				

\usepackage{amsmath}
\usepackage{amssymb}
\usepackage{amsthm}

\usepackage{mathtools}

\usepackage{feynmf}
\usepackage{marvosym}

\usepackage{import}

\newtheoremstyle{named}{}{}{\itshape}{}{\bfseries}{.}{.5em}{#3}
\theoremstyle{named}

\newcommand*\xoverline[2][0.75]{%
    \sbox{\myboxA}{$\m@th#2$}%
    \setbox\myboxB\null
    \ht\myboxB=\ht\myboxA%
    \dp\myboxB=\dp\myboxA%
    \wd\myboxB=#1\wd\myboxA
    \sbox\myboxB{$\m@th\overline{\copy\myboxB}$}
    \setlength\mylenA{\the\wd\myboxA}
    \addtolength\mylenA{-\the\wd\myboxB}%
    \ifdim\wd\myboxB<\wd\myboxA%
       \rlap{\hskip 0.5\mylenA\usebox\myboxB}{\usebox\myboxA}%
    \else
        \hskip -0.5\mylenA\rlap{\usebox\myboxA}{\hskip 0.5\mylenA\usebox\myboxB}%
    \fi}
\makeatother

\newcommand{\Mpl}{M_{\textrm{Pl}}}

\definecolor{cobalt}{RGB}{44, 98, 120}
\definecolor{celadon}{rgb}{0.67, 0.88, 0.69}
\definecolor{dm}{cmyk}{.20, 0, .30, 0}
\definecolor{burgundy}{rgb}{0.5, 0.0, 0.13}
\definecolor{plotBlue}{RGB}{94, 130, 181}

\newcommand{\rmd}{\textrm{d}}
\def\be{\begin{equation}}
\def\ee{\end{equation}}
\def\bea{\begin{eqnarray}}
\def\eea{\end{eqnarray}}

\hypersetup{
  colorlinks,
  citecolor=Violet,
  linkcolor=cobalt,
  urlcolor=Blue}

\newif\iffastcompile

\fastcompilefalse

\iffastcompile
\newcommand{\js}[1]{}
\newcommand{\jsi}[1]{}
\newcommand{\cl}[1]{}
\newcommand{\lm}[1]{}
\else
\newcommand{\js}[1]{\todo[color=cobalt!30,size=\scriptsize, bordercolor=cobalt!30]{JS: #1}}
\newcommand{\jsi}[1]{\todo[color=cobalt!30,size=\scriptsize, bordercolor=cobalt!30, inline]{JS: #1}}
\newcommand{\cl}[1]{\todo[color=burgundy!30, size=\scriptsize, bordercolor=burgundy!30]{CL: #1}}
\newcommand{\lm}[1]{\todo[color=dm!90, size=\scriptsize, bordercolor=dm!90]{LM: #1}}
\fi

\ProvideTextCommandDefault{\Dbar}{%
\leavevmode\lower.5ex\rlap{\hskip-.07em\accent"16}D%
}

\begin{document}
	\newcommand{\main}{.}
\begin{titlepage}

\setcounter{page}{1} \baselineskip=15.5pt \thispagestyle{empty}

\bigskip\

\vspace{1.05cm}
\begin{center}
{\fontsize{20}{28} \bfseries Higher-Group Symmetries \\ \vspace{0.3cm} and Weak Gravity Conjecture Mixing}

 \end{center}
\vspace{1cm}

\begin{center}
\scalebox{0.95}[0.95]{{\fontsize{14}{30}\selectfont Sami Kaya and Tom Rudelius}}
\end{center}

\begin{center}
\vspace{0.25 cm}
{Department of Physics, University of California, Berkeley, CA 94720 USA}\\

\vspace{0.25cm}

\end{center}

\vspace{1cm}
\noindent 

In four-dimensional axion electrodynamics, a Chern-Simons coupling of the form $\theta  F \wedge F$ leads to a higher-group global symmetry between background gauge fields. At the same time, such a Chern-Simons coupling leads to a mixing between the Weak Gravity Conjectures for the axion and the gauge field, so that the charged excitations of a Weak Gravity Conjecture-satisfying axion string will also satisfy the Weak Gravity Conjecture for the gauge field. In this paper, we argue that these higher-group symmetries and this phenomenon of Weak Gravity Conjecture mixing are related to one another. We show that this relationship extends to supergravities in 5, 6, 7, 8, 9, and 10 dimensions, so higher-dimensional supergravity is endowed with precisely the structure needed to ensure consistency with emergent higher-group symmetries and with the Weak Gravity Conjecture. We further argue that a similar mixing of Weak Gravity Conjectures can occur in two-term Chern-Simons theories or in theories with kinetic mixing, though the connection with higher-group symmetries here is more tenuous, and accordingly the constraints on effective field theory are not as sharp.

 \vspace{1.1cm}

\bigskip
\noindent\today

\end{titlepage}
\setcounter{tocdepth}{2}
\tableofcontents

\section{Introduction}\label{INTRO}

Recent years have seen a revolution in our understanding of symmetries in quantum field theory. The familiar notion of a global symmetry, it turns out, is merely one example of a rich zoo of related phenomena, which go under names like higher-form symmetries \cite{Gaiotto:2014kfa}, exotic symmetries \cite{Seiberg:2019vrp}, cobordisms \cite{McNamara:2019rup}, non-invertible symmetries, and higher-group symmetries \cite{Sharpe:2015mja, Cordova:2018cvg, Benini:2018reh}.

These generalizations of global symmetries have also played an increasingly important role in the quest to understand and characterize field theories in the quantum gravity landscape. 
It has long been argued that exact global symmetries are forbidden in quantum gravity \cite{Hawking:1974sw, Zeldovich:1976vq, Zeldovich:1977be, Banks:1988yz, Giddings:1988cx,Abbott:1989jw,Coleman:1989zu,Kallosh:1995hi, Witten:1998qj, Banks:2010zn, Harlow:2018tng, Harlow:2020bee, Chen:2020ojn,Hsin:2020mfa, Yonekura:2020ino}, and likely this extends to other generalizations of symmetries as well \cite{Harlow:2018tng,  Rudelius:2020orz, McNamara:2019rup}. 
This means that any global symmetry in an effective field theory must be removed when the theory is coupled to gravity, either through breaking or gauging the symmetry. This in turn is related to many familiar phenomena in quantum field theory and quantum gravity \cite{Heidenreich:2020pkc}, and to the Completeness Hypothesis \cite{polchinski:2003bq, Harlow:2018tng, Rudelius:2020orz, Heidenreich:2021tna, Casini:2021zgr}, which holds that any EFT in the landscape should have states in every representation of its gauge group.

While exact global symmetries may be forbidden in quantum gravity, there is no law forbidding emergent and/or approximate global symmetries at low energies. Emergent global symmetries--and their breaking at high energies--has interesting connections with black hole physics \cite{Nomura:2019qps, Fichet:2019ugl} and the Weak Gravity Conjecture (WGC) \cite{CordovaOhmoriRudelius, Daus:2020vtf}, a conjecture which demands the existence of light particles charged under a gauge group.\footnote{For more on the WGC, see the original paper \cite{Arkanihamed:2006dz} or the recent review article \cite{Harlow:2022gzl}.}

There is likewise no restriction against emergent and/or approximate higher-form symmetries, non-invertible symmetries, higher-group symmetries, and the like. Recently, Brennan and C\'ordova \cite{Brennan:2020ehu} have argued that emergent higher-group global symmetries impose interesting constraints on effective field theory. More explicitly, in a higher-group global symmetry, two higher-form global symmetries are mixed up with one another, and one of these two symmetries is subordinate to the other in the sense that it cannot exist while the other is broken. This means that the superior symmetry must emerge at an energy scale at or above the emergence scale of the subordinate symmetry, which leads to meaningful constraints on field theory parameters.

A particularly notable example of a field theory with an emergent higher-group symmetry is axion electrodynamics, which involves a periodic scalar field (axion) $\theta$ coupled to electromagnetism via a $\theta F \wedge F$ coupling. Here, a higher-group symmetry emerges in which the superior symmetry has Noether current $\rmd \theta$, while the subordinate symmetry shifts the gauge field $A$ by a flat connection. Consistency of the theory then requires that the latter symmetry must emerge below the scale of the former.

This same theory of axion electrodynamics was also recently investigated in relation to the WGC in \cite{Heidenreich:2021yda}. In particular, if we denote the axion decay constant by $f_\theta$ and assume (a) an ``axion string'' (i.e., a string around which $\theta$ undergoes monodromy, $\theta \rightarrow \theta+ 2 \pi$) of tension
\begin{equation}
T \lesssim f_\theta \Mpl,
\label{Teq}
\end{equation} 
(b) an instanton of action
\begin{equation}
S \lesssim \Mpl / f_\theta\,,
\label{Seq}
\end{equation}
and (c) a relationship $S \sim 1/e^2$ between the instanton action and the electric gauge coupling $e$,
then \cite{Heidenreich:2021yda} argued that the string excitations of the axion string should carry charge under the gauge field $A$ and have a mass of order
\begin{equation}
m \lesssim e \Mpl\,,
\label{meq}
\end{equation}
Here, \eqref{Teq} is simply the WGC for axion strings, \eqref{Seq} is the WGC for instantons, and \eqref{meq} is the WGC for charged particles. From this, we see that these distinct versions of the WGC are ``mixed up'' with one another in the presence of the $\theta F \wedge F$ coupling.

In this paper, we will argue that within the context of axion electrodynamics, the mixing of global symmetries into a higher-group structure and the mixing of WGCs are are in fact related to one another, thereby unifying the recent observations of \cite{Brennan:2020ehu} and \cite{Heidenreich:2021yda}. We will also show that these 4d results can be generalized to a wide array of higher dimensions, as the structure of supergravity in dimensions 5-10 concurrently produces both emergent higher-group symmetries and WGC mixing amidst couplings of the form $C_{d-4} \wedge F \wedge F$.

We will further generalize our analysis to theories with other couplings between different gauge fields. Though the connection with higher-group symmetries is not as sharp in these contexts, we will see that a similar sort of WGC mixing occurs in BF theories and in theories with kinetic mixing. The former plays a role in the abelian Higgs mechanism and axion monodromy inflation \cite{McAllister:2008hb,Silverstein:2008sg}, while the latter plays a role in decay constant alignment inflation \cite{kim:2004rp}.

The remainder of this paper is structured as follows: in \S\ref{HIGHER}, we review higher-form and higher-group global symmetries. In \S\ref{Strings}, we review higher-group symmetries and WGC mixing in axion electrodynamics, then generalize these observations to supergravites in 5, 6, 7, 8, 9, and 10 dimensions. In \S\ref{BF}, we discuss the mixing of higher-form symmetries and WGCs in BF theories, and in \S\ref{ALIGNMENT} we do the same for theories with kinetic alignment.

In what follows, we use capital letters $A, B, C, ...$ to denote dynamical gauge fields and lower case letters $a, b, c, ...$ to denote background gauge fields. Form degrees are labeled by subscripts, so e.g. $A_1$ represents a dynamical 1-form gauge field, while $h_3$ represents a background 3-form field strength. Subscripts will at times be omitted when they are clear from context. The signature of spacetime is taken to be $(-, +, ..., +)$, and supergravity actions are always written in Einstein frame. We work in units with $8 \pi G = 1$, though at times we will restore factors of the reduced Planck mass for sake of exposition.

\section{Higher-Group Global Symmetries}\label{HIGHER}

In this section, we review relevant aspects of higher-group global symmetries. For further reading on the topic, consult \cite{Cordova:2018cvg, Benini:2018reh}.

First, we introduce the related notion of a higher-form global symmetry \cite{Gaiotto:2014kfa}. A $p$-form global symmetry is a symmetry whose charged operators are supported on manifolds of dimension $p$. An ordinary global symmetry is therefore a 0-form global symmetry: the charged operators are local operators, such as a field which rotates by a phase.

A $p$-form global symmetry is associated with a group, $G$.\footnote{Recently, there has been significant interest in \emph{non-invertible} global symmetries, which are not associated with a group, but rather a more complicated fusion algebra. See e.g. \cite{Bhardwaj:2017xup, Chang:2018iay, Thorngren:2019iar, Rudelius:2020orz, Heidenreich:2021tna, Sharpe:2021srf, Koide:2021zxj, Choi:2021kmx} for more on this topic.} For an ordinary symmetry, i.e. $p=0$, $G$ may be either abelian or nonabelian, but for a higher-form symmetry $p>0$, $G$ is necessarily abelian.

$G$ may be either discrete or continuous. In the continuous case, under reasonable circumstance, there will be an associated Noether current, which is a closed $d-p-1$-form, i.e., 
\begin{equation}
\rmd J_{d-p-1} = 0\,.
\end{equation}
We may couple a $p$-form global symmetry to a $(p+1)$-form background gauge field $A_{p+1}$, which will transform under a background gauge transformation as 
\begin{equation}
A_{p+1} \rightarrow A_{p+1} + \rmd \Lambda_{p}\,.
\label{Atransform}
\end{equation} 

With this lightning review of higher-form global symmetries, we are now in a position to introduce the notion of a higher-group global symmetry. Suppose we have a collection of $p_k$-form global symmetries with background gauge fields $A_{p_k+1}^{(k)}$. Each such field will transform under its own symmetry transformation as in \eqref{Atransform}, but it may also transform under background gauge transformations of the other gauge fields as
\begin{equation}
A_{p_k+1}^{(k)} \rightarrow A_{p_k+1}^{(k)} + \rmd \Lambda_{p_k}^{(k)} + \sum_i \alpha_{p_k + 1 - p_i}(A_{p_j}^{(j)}) \Lambda_{p_i} + \text{ Schwinger terms,}
\label{highergroup}
\end{equation}
where $p_i < p_k +1$. Here, $\alpha_{p_k + 1 - p_i}(A_{p_j}^{(j)})$ is a $(p_k + 1 - p_i)$-form consisting of products of the background gauge fields $A_{p_j}^{(j)}$, and Schwinger terms are terms that are nonlinear in the gauge parameters. This collection of gauge fields is then referred to as a higher-group global symmetry, and specifically it is called a $p$-group if $p = \text{max}(p_k+1)$ is the largest form degree of a background gauge field transforming nontrivially under the higher-group structure.

In what follows, we will see examples of higher-form symmetries, higher-group symmetries, and further generalizations thereof.

\section{Magnetic Strings and $C FF$ Couplings}\label{Strings}

In this section, we discuss WGC mixing and higher-group structures in theories with $C_{d-4} \wedge F_2 \wedge F_2$ Chern-Simons couplings. This WGC mixing was previously discussed in 4d and 5d in \cite{Heidenreich:2021yda}, but here we extend the analysis to higher dimensions and elaborate on connections to higher-group symmetries.

We begin with a discussion of higher-group symmetries in such theories, generalizing the work of \cite{Brennan:2020ehu} (see also \cite{Tanizaki:2019rbk}). Our starting point is the action
\begin{equation}
S = \int - \frac{1}{2 g_A^2} F_2 \wedge \star F_2 - \frac{1}{2 g_C^2} G_{d-3} \wedge \star G_{d-3} -\frac{K }{8 \pi^2} C_{d-4} \wedge F_2 \wedge F_2 \,,
\label{CFF}
\end{equation}
where $F_2 = \rmd A_1$, $G_{d-3} = \rmd C_{d-4}$ and $K$ is an integer.

We may couple this system to background gauge fields as follows:
\begin{align}
S &= \int_X \Big[ - \frac{1}{2 g_A^2} (F_2 - a_{2}^{(e)}) \wedge \star (F_2 - a_{2}^{(e)}) - \frac{1}{2 g_C^2} (G_{d-3} -   c_{d-3}^{(e)}) \wedge \star (G_{d-3} -   c_{d-3}^{(e)})  \nonumber \\
&- \frac{1}{2 \pi} A_1 \wedge \rmd a_{d-2}^{(m)}  -  \frac{1}{2 \pi} C_{d-4} \wedge \rmd c_{3}^{(m)} \Big]  +\frac{K }{8 \pi^2}  \int_Y (G_{d-3} -   c_{d-3}^{(e)}) \wedge (F_2 - a_{2}^{(e)}) \wedge (F_2 - a_{2}^{(e)}) \,.
\end{align}
Here, $X$ is the $d$-dimensional spacetime, $Y$ is a $d+1$-manifold whose boundary is $X$, and the electric background gauge fields satisfy
\begin{equation}
K a_{2}^{(e)} = \rmd \lambda_1^{(a)}  \,,~~~~ K c_{d-3}^{(e)} = \rmd \lambda_{d-2}^{(c)}\,,
\end{equation}
where $\lambda_1^a$ and $\lambda_{d-2}^c$ are gauge parameters with periods $\rmd \lambda$ valued in $2 \pi \mathbb{Z}$, reflecting the fact that the Chern-Simons coupling breaks the electric global symmetries to a $\mathbb{Z}_K$ subgroup. The ``electric'' symmetries, indicated by an $e$ superscript, are so named because the Wilson line/surface operators of $A_1$ and $C_{d-4}$ are charged under these symmetries, whereas 't Hooft operators for $A_1$ and $C_{d-4}$ are charged under the ``magnetic'' symmetries, which are labeled by an $m$ subscript.

The gauge transformations for the electric background fields are standard:
\begin{equation}
a_{2}^{(e)} \rightarrow  a_{2}^{(e)} + \rmd \Lambda_1^{(a, e)}\,,~~~c_{d-3}^{(e)} \rightarrow  c_{d-3}^{(e)} + \rmd \Lambda_{d-4}^{(c,e)}\,,
\end{equation}
but the gauge transformations for the magnetic background fields are mixed up into a higher-group structure:
\begin{align}
a_{d-2}^{(m)} &\rightarrow  a_{d-2}^{(m)} + \rmd \Lambda_{d-3}^{(a, m)} - \frac{K}{2\pi} \Lambda_1^{(a,e)} \wedge c_{d-3}^{(e)} - \frac{K}{2 \pi} \Lambda_{d-4}^{(c, e)} \wedge a_2^{(e)} - \frac{K}{2 \pi} \Lambda_1^{(a,e)} \wedge \rmd \Lambda_{d-4}^{(c,e)}  \\
c_{3}^{(m)} &\rightarrow  c_{3}^{(m)} + \rmd \Lambda_{2}^{(c,m)} - \frac{K}{2 \pi} a_2^{(e)} \wedge \rmd \Lambda_1^{(a,e)} - \frac{K}{4 \pi} \Lambda_1^{(a, e)} \wedge \rmd \Lambda_1^{(a, e)} \,,
\label{mmixing}
\end{align}
The gauge-invariant field strengths are given by
\begin{align}
f_{d-1}^{(m)} &= \rmd a_{d-2}^{(m)} + \frac{K}{2 \pi} c_{d-3}^{(e)} \wedge a_2^{(e)}  \\
g_{4}^{(m)} &= \rmd c_{3}^{(m)} + \frac{K}{4 \pi} a_{2}^{(e)} \wedge a_2^{(e)} \, . \label{eqg}
\end{align}
The appearance of $\Lambda_1^{a,e}$ in the transformation of $c_3^{(m)}$ underscores an important fact: in the presence of the Chern-Simons coupling, electric charge of $A_1$ can dissolve in the worldsheet of the string charged magnetically under $C_{d-4}$, so the string states of this string may carry $A_1$ charge. We will see that this plays an important role in WGC mixing in the examples that follow.

Suppose now that we have a renormalization group trajectory which flows to our theory \eqref{CFF}, so that a $\mathbb{Z}_K$ electric 1-form symmetry emerges at low energies. As noted by Brennan and C\'ordova \cite{Brennan:2020ehu},\footnote{The analysis of \cite{Brennan:2020ehu} was specific to four dimensions, but it may be straightforwardly generalized to arbitrary dimensions.} \eqref{eqg} implies any background for $a_2^{(e)}$ necessarily turns on a background for $c_3^{(m)}$, which means that the 2-form magnetic symmetry must also be a good symmetry of the theory. If $\Lambda_{\text{screen}}$ is the energy scale below which the 1-form symmetry emerges and $\Lambda_{\text{string}}$ is the energy scale below which the 2-form symmetry emerges, we then have the constraint
\begin{equation}
\Lambda_{\text{screen}} \lesssim \Lambda_{\text{string}}\,.
\end{equation}

From here, we observe that the electric 1-form symmetry is broken by charged particles (see e.g. \cite{Rudelius:2020orz, Heidenreich:2021tna}), which means that the mass $m$ of the lightest charged particle necessarily satisfies 
\begin{equation}
m \lesssim \Lambda_{\text{screen}}\, .
\label{mscreenb}
\end{equation}
Furthermore, for a string of tension $T$, we have \cite{Brennan:2020ehu}
\begin{equation}
\Lambda_{\text{string}} \lesssim \sqrt{T} \,.
\label{Lstringb}
\end{equation}
In other words, the 2-form symmetry breaks down at a scale no larger than the string scale $M_{\textrm{string}} := \sqrt{2 \pi T}$.

Next, we use the fact that in the presence of a $C_{d-4} \wedge F_2 \wedge F_2$ coupling (or a $C_{d-4} \wedge \textrm{Tr} (F_2 \wedge F_2)$ coupling, if the gauge group is nonabelian), we expect instantons with tension 
\begin{equation}
T_{\textrm{inst}} \sim \frac{8 \pi^2}{g_A^2}\,.
\label{insteq}
\end{equation}
Here, an ``instanton'' is an object of codimension 4, which is charged under $C_{d-4}$. For a nonabelian gauge group, these are the usual Yang-Mills instantons, whereas for an abelian group these instantons arise from monopole loops \cite{Fan:2021ntg}.\footnote{The monopole loop calculation of \cite{Fan:2021ntg} was also carried out in four dimensions, but it can be extended to higher dimensions straightforwardly, as the $d-4$ dimensions of the instanton worldvolume simply come along for the ride. We thank Matthew Reece for pointing this out to us.}

Finally, we assume that these codimension-4 instantons satisfy the WGC for the gauge field $C_{d-4}$:
\begin{equation}
\frac{8 \pi^2}{g_A^2} \sim T_{\textrm{inst}} \lesssim  g_C M_{\text{Pl}; d}^{(d-2)/2} \,,
\label{gscaling}
\end{equation}
and we assume that there exists a charged string of tension $T_{\textrm{string}}$ which satisfies the magnetic WGC for $C_{d-4}$:
\begin{equation}
T_{\textrm{string}} \lesssim g_C^{-1} M_{\text{Pl}; d}^{(d-2)/2} \,.
\end{equation}

Together with \eqref{insteq}, these bounds imply
\begin{equation}
T_{\textrm{string}} \lesssim g_C^{-1} M_{\text{Pl}; d}^{(d-2)/2} \lesssim g_A^2 M_{\text{Pl}; d}^{d-2}\,.
\end{equation}
Combining this with the consistency conditions for the higher-group structure \eqref{mscreenb}, \eqref{Lstringb}, we conclude that the mass of the lightest charged particle satisfies
\begin{equation}
m \lesssim  \Lambda_{\text{screen}} \lesssim \Lambda_{\text{string}} \lesssim \sqrt{T_{\textrm{string}}} \lesssim g_C^{-1/2} M_{\text{Pl}; d}^{(d-2)/4} \lesssim g_A M_{\text{Pl}; d}^{(d-2)/2}\,,
\label{scalebound}
\end{equation}
which implies (up to $O(1)$ factors) the WGC for the gauge field $A_1$, $m \lesssim g_A M_{\text{Pl}; d}^{(d-2)/2}$.

One can alternatively argue for the WGC in these contexts on the basis of anomaly inflow on the string worldsheet, following \cite{Heidenreich:2021yda}, which implies that the string states of mass $M_{\textrm{string}}$ must carry electric charge, which together with the bound $M_{\textrm{string}} \lesssim g_A M_{\text{Pl}; d}^{(d-2)/2}$ implies that the WGC will be satisfied for $A_1$. Of course, these are two sides of the same coin. 

We can say even more if the charged string is a fundamental string, i.e., if the core of the string probes physics in the deep ultraviolet \cite{Reece:2018zvv}. In this case, there is a whole tower of superextremal string excitations, i.e., string excitations which satisfy the WGC bound. This means that in addition to the WGC, the tower and sublattice WGC \cite{Heidenreich:2015nta, Heidenreich:2016aqi, Andriolo:2018lvp} are satisfied for the gauge field $A_1$ as well, and furthermore the tower of superextremal charged particles stipulated by these conjectures is a stringy one. In four dimensions, this has interesting phenomenological consequences \cite{Heidenreich:2021yda}.

The upshot of all of this is that, in a sense, the electric WGC for $A_1$ and the magnetic WGC for $C_{d-4}$ have been ``mixed up'' with one another in the presence of the $C_{d-4} \wedge F_2 \wedge F_2$ coupling: the excitations of a string that satisfies the magnetic WGC for $C_{d-4}$ will satisfy the electric WGC for $A_1$. Meanwhile, the Chern-Simons coupling also implies that the electric 1-form symmetry for $A_1$ and the magnetic 2-form symmetry for $C_{d-4}$ are tied up into a higher-group structure \eqref{mmixing}, and the consistency condition \eqref{scalebound} for this higher-group structure implies the WGC for $A_1$. The mixing of higher-form symmetries into a higher group structure is thus intimately connected with the mixing of Weak Gravity Conjectures for the gauge fields $A_1$ and $C_{d-4}$.

It is worth noting, however, that this WGC mixing persists when $|K| = 1$ even though the higher-group structure disappears. This suggests that we should not view the higher-group symmetry as the fundamental principle underlying the WGC mixing, but rather the Chern-Simons term itself is the fundamental cause of both the WGC mixing and the higher-group symmetry. We will see more evidence for this perspective in the following section, where a version of WGC mixing exists in the presence of a Chern-Simons term even though a higher-group structure is absent. Nonetheless, the fact that the higher-group consistency bound \eqref{scalebound} leads to WGC mixing for $|K| > 1$ suggests that these phenomena are closely tied to one another, even if one does not always underlie the other. 

Finally, it is worth noting that consistency of the higher-group structure places an upper bound \eqref{scalebound} on the mass of a superextremal charged particle that is well below the Planck scale in a weakly coupled gauge theory. This suggests that the WGC should be satisfied not only by black holes with subleading corrections to their charge-to-mass ratios \cite{Kats:2006xp}, but also by light particles in the effective field theory. In the case of a fundamental string, this suggests that the sublattice WGC should be satisfied by a sublattice of superextremal particles that is not too sparse.

In the remainder of this section, we will show how the Chern-Simons couplings studied here lead to WGC mixing in diverse dimensions, as expected from higher-group consistency conditions and/or anomaly inflow on the string worldsheet. In particular, we will see that the relation $g_C^{-1} \sim g_A^2$ is imposed by supersymmetry in dimensions 5-10. By \eqref{scalebound}, this means that the magnetic WGC for $C_{d-4}$ implies the WGC for $A_1$. Thus, WGC mixing is built into the structure of supergravity.

\subsection{Review: 4d axion strings}\label{ssec:4d}

We begin by reviewing the 4d story, which may have important implications for phenomenology. This story was originally presented in greater detail in \cite{Heidenreich:2021yda}.

In four dimensions, the action \eqref{CFF} takes the form:
\begin{equation}
S = \int - \frac{1}{2 g_A^2} F_2 \wedge \star F_2 - \frac{f_\theta^2}{2} \rmd \theta \wedge \star \rmd \theta + \frac{ K }{8 \pi^2} \theta \wedge F_2 \wedge F_2 \,,
\label{CFF4d}
\end{equation}
where $f_{\theta}$ is the axion decay constant. The axion WGC for the field $\theta$ implies
\begin{equation}
f_\theta \lesssim \frac{\Mpl}{S}\,,
\label{fbound}
\end{equation}
where $S$ is the instanton action. The WGC for strings implies
\begin{equation}
T \lesssim f_\theta \Mpl\,,
\end{equation}
where $T$ is the tension of the axion string charged magnetically under $\theta$.

Here, we make one additional assumption, which is that the instanton action takes the form
\begin{equation}
S = \frac{8 \pi^2}{g_A^2}\,.
\label{Sexp}
\end{equation}
This form of the instanton action is most familiar from QCD, but it is also valid for abelian gauge theories, with instantons coming in the form of monopole loops with dyonic winding~\cite{Fan:2021ntg}.

Together, \eqref{fbound}-\eqref{Sexp} imply a bound
\begin{equation}
T \lesssim \frac{g_A^2}{4 \pi} \Mpl^2\,.
\end{equation}
This in turn places a bound on the string scale of the form
\begin{equation}
M_{\textrm{string}} := \sqrt{2 \pi T} \lesssim g_A \Mpl\,.
\end{equation}
Because of the $\theta F \wedge F$ coupling, the excitations of this string will carry electric charge under the gauge field $A$. This follows from anomaly inflow on the string worldsheet \cite{Callan:1984sa}. Such an excitation has string scale mass:
\begin{equation}
m \sim M_{\textrm{string}} \lesssim g_A \Mpl\,,
\end{equation}
which is simply the statement that this charged string excitation satisfies the WGC, up to $O(1)$ factors. We learn that the WGC for particles is guaranteed by the WGC for axions and axion strings in the presence of the $\theta F \wedge F$ Chern-Simons coupling, assuming the instanton action takes the form \eqref{Sexp}. Note that the inequalities in \eqref{scalebound} are also obeyed, thanks to the superextremal charged string excitations, as they must be for consistency of the higher-group structure when $|K| > 1$.

Even more can be said if the axion in question is a fundamental axion, i.e., core of the axion string probes physics in the deep ultraviolet \cite{Reece:2018zvv}. In this case, there is a whole tower of superextremal string excitations, so not only is the WGC satisfied, but the tower and sublattice WGC \cite{Heidenreich:2015nta, Heidenreich:2016aqi, Andriolo:2018lvp} are satisfied for the gauge field $A_1$ as well.

This is our first example of WGC mixing in the presence of Chern-Simons terms (and their accompanying higher-group structure): the WGC for particles follows from the electric and magnetic WGCs for the axion in the presence of the $\theta F \wedge F$ term. Certainly, this is not the case in the absence of such a $\theta F \wedge F$ term: here, the string excitations will not carry electric charge, and the expression \eqref{Sexp} will be invalid.

An illustrative example in which the $\theta F \wedge F$ coupling is not present occurs in the compactification of minimal 5d supergravity on a circle. In the presence of a Chern-Simons term $A_1 \wedge F_2 \wedge F_2$ in 5d, one finds that the resulting 4d theory has an axion string with string scale \cite{Heidenreich:2021yda}
\begin{equation}
M_{\textrm{string}} = \sqrt{2 \pi T} \sim e_{\textrm{KK}}^{1/3} \Mpl \sim M_{\textrm{Pl};5}\,,
\end{equation}
where $e_{\textrm{KK}} = \sqrt{2}/(R \Mpl)$ is the gauge coupling of the KK gauge field. The fact that the string scale is the 5d Planck scale indicates that the theory is secretly a 5d gravity theory, so effective field theory breaks down at the 5d Planck scale rather than the 4d Planck scale, the latter of which is much larger at large radius. The fact that the string scale is NOT given by $ e_{\textrm{KK}}  \Mpl$ is an indication of the absence of a Chern-Simons coupling $\theta F^{(\textrm{KK})} \wedge F^{(\textrm{KK})}$ between the KK photon and the axion, so the preceding argument does not apply to this scenario.

An important corollary to the above analysis is that we expect effective field theory to break down at the string scale $M_{\textrm{string}} \sim g_A \Mpl$ in the presence of the Chern-Simons coupling $\theta F \wedge F$, whereas effective field theory may remain valid up to the higher scale $g_A^{1/3} \Mpl$ in the absence of such a Chern-Simons coupling. This has important phenomenological consequences for effective field theories requiring both tiny gauge couplings and high energy scales \cite{Heidenreich:2017sim, Heidenreich:2021yda}.

\subsection{Review: strings in 5d supergravity}

We will now see that a very similar story plays out in 5d supergravity: the presence of a Chern-Simons coupling $C \wedge F \wedge F$ between two 1-form gauge fields leads to WGC mixing, thereby preserving a higher-group structure between background gauge fields. A more detailed presentation can be found in \cite{Heidenreich:2021yda, BPSstrings}.

We suppose we have a 5d supergravity theory with exactly two 1-form gauge fields, $A_1$ and $C_1$. The relevant part of the action takes the form
\begin{equation}
S \supset \int -\frac{1}{2 g_A^2} F_2 \wedge \star F_2 - \frac{1}{2 g_C^2} G_{2} \wedge \star G_{2} - \frac{1}{2 g_{AC}^2 } (F_2 \wedge \star G_2 + G_2 \wedge \star F_2 )  + \frac{K }{8 \pi^2} C_1\wedge F_2 \wedge F_2  \,.
\end{equation}
Due to the supersymmetry, the Chern-Simons coupling $K$ and the gauge couplings $g_A$, $g_C$, and $g_{AC}$ are correlated: all of them can be computed in terms of the cubic prepotential. We defer the reader to \cite{Heidenreich:2021yda} for the details, and instead simply quote the main results.

If $K \neq 0$, then in the limit $g_A \rightarrow 0$, then the structure of the 5d prepotential gives
\begin{equation}
g_C \sim g_A^{-2} M_{\textrm{Pl}; 5}^{-3/2}\,,~~~\frac{1}{g_{AC}} = 0 \,.
\label{CA1}
\end{equation}
The latter equality implies that there is no mixing whatsoever between the two gauge fields. The former relation implies that the charge of a string charged magnetically under $C_1$ vanishes in the $g_A \rightarrow 0$ limit. Excitations of this string will carry charge under $A_1$, by the same anomaly inflow argument as in 4d above.
If we further suppose that the string satisfies the WGC:
\begin{equation}
T \lesssim g_C^{-1} M_{\textrm{Pl}; 5}^{3/2}\,,
\end{equation}
then we conclude from \eqref{CA1} that its lightest string excitations will have a mass $m$ of the form
\begin{equation}
m \sim M_{\textrm{string}} = \sqrt{2 \pi T} \lesssim g_A M_{\textrm{Pl}; 5}^{3/2}\,.
\end{equation}
From this inequality, we learn that these string excitations will be superextremal and satisfy the WGC. If the core of the string probes the deep ultraviolet, we expect an entire tower of superextremal string modes, satisfying the tower/sublattice WGC for $A_1$. This is our second example of WGC mixing in the presence of Chern-Simons terms: the WGC for $A_1$ follows from the magnetic WGC for the string charged under $B_1$, under the additional constraints of supersymmetry. Note also that the WGC bound for $A_1$ is consistent with the inequality \eqref{scalebound}, as required by the higher-group structure for $|K| > 1$.

On the other hand, if $K = 0$, so there is no Chern-Simons coupling, then one can show:
\begin{equation}
g_C \sim g_A^{-1/2} M_{\textrm{Pl}; 5}^{-3/4}\,,~~~\frac{1}{g_{AC}} = O(g_A^{5/4} M_{\textrm{Pl};5}^{9/8}) \,.
\label{CA2}
\end{equation} 
The latter equality implies that any mixing between the two gauge fields is heavily suppressed, and therefore insignificant, in the $g_A \rightarrow 0$ limit. The former relation implies that the charge of a string charged magnetically under $C_1$ vanishes in the $g_A \rightarrow 0$ limit. However, assuming this string satisfies the WGC, its string scale takes the form
\begin{equation}
M_{\textrm{string}} = \sqrt{2 \pi T} \lesssim g_A^{1/4} M_{\textrm{Pl}; 5}^{9/8}\,.
\end{equation}
This one-quarter scaling parallels the one-third scaling seen in the Kaluza-Klein example in four dimensions above: if $A_1$ is a Kaluza-Klein gauge field, so that $g_A \sim 1/(R M_{\textrm{Pl};5}^{3/2})$, then the string scale here is nothing but the 6d Planck scale, $M_{\textrm{string}} \sim M_{\textrm{Pl};6} = R^{-1/4} M_{\textrm{Pl};5}^{3/4}$. We expect, therefore, that $K=0$ whenever $A_1$ is a Kaluza-Klein gauge field, and $g_A \rightarrow 0$ is a decompactification limit.

\subsection{Strings in 6d supergravity}

We will now argue that a similar phenomenon occurs in 6d supergravity.  To begin, we review relevant aspects of 6d supergravity coupled to abelian gauge fields, following \cite{Riccioni:1999xq} (see also \cite{Sagnotti:1992qw, Riccioni:2001bg}). 

A generic 6d supergravity features one supergravity multiplet,  $n_T$ tensor multiplets, and $n_V$ vector multiplets. It may also include hypermultiplets, but we can safely ignore these multiplets, as they do not affect the kinetic terms of the gauge fields or the Chern-Simons couplings between them. The supergravity multiplet includes the metric and an anti-self-dual 2-form gauge field, but no scalar field. A tensor multiplet features one self-dual 2-form gauge field and a scalar field. A vector multiplet features a 1-form gauge field, but no scalar field.

The upshot of this is that a theory with $n_T$ tensor multiplets will have $n_T+1$ 2-form gauge fields and $n_T$ scalar fields. (Again, we can safely ignore scalar fields in the hypermultiplets.) The scalar fields parametrize the coset $SO(1, n_T)/SO(n_T)$, which is also known as the tensor multiplet moduli space. We may describe these in terms of the $SO(1, n_T)$ matrix
\begin{equation}
V = \left( \begin{array}{c}
v_r \\
x_r^M
\end{array}  \right)\, ,
\end{equation}
where $r=0,1 , ..., n_T$ and $M = 1, 2, ..., n_T$. These are subject to the conditions
\begin{equation}
v^r v_r = 1\,,~~~- v_r v_s + x^M_r x^M_s = \eta_{rs}\,,~~~v^r x_r^M =0\,,
\label{6dcond}
\end{equation}
where here, repeated indices are summed, and $r$ and $s$ indices are raised and lowered via the metric $\eta_{rs} = \text{diag}(-1, 1, 1, ..., 1)$. Thus, for instance, $v^0 = -v_0$, $v^1 =  v_1$.

The gauge kinetic matrix for the tensor fields is given by
\begin{equation}
G_{rs} = v_r v_s + x_r^M x_s^M\,.
\end{equation}

Then, the relevant part of the action is given by 
\begin{align}
S =  \int  - \frac{1}{2} G_{rs} H_3^{r } \wedge \star H_{3}^s +\frac{1}{4} \rmd v^r \wedge \star \rmd v_r + \frac{1}{2} v_r c^{rab} F^a \wedge \star F^{b} + \frac{1}{2} c_r^{ab} B_2^r \wedge F_2^a  \wedge F_{2}^b\,.
\label{action}
\end{align}
Here, $a, b = 1,...,n_V$ run over the (abelian) vector multiplets, and these indices can be raised and lowered freely. The $c_r^{ab}$ are constants.

\subsubsection{One tensor multiplet}

In the particular case $n_T = 1$, there is just one scalar field, and the equations for the scalar field matrix $V$ can be solved simply as
\begin{equation}
V = \left( \begin{array}{cc}
\cosh(\phi) & \sinh (\phi) \\
\sinh (\phi) & \cosh (\phi)
\end{array}  \right)\, .
\end{equation}
We then have
\begin{equation}
 \frac{1}{4} \rmd v^r \wedge \star \rmd v_r = - \frac{1}{4} \rmd \phi \wedge \star \rmd \phi\,
\end{equation}
for the scalar kinetic term. We further have 
\begin{equation}
G_{00} =G_{11} =  \cosh(2\phi)  \,,~~G_{01} = G_{10} =  \sinh (2 \phi) \,
\end{equation}
for the 2-form gauge kinetic term. This metric has eigenvalues and eigenvectors given respectively by
\begin{equation}
\lambda_1 = e^{-2 \phi} \,,~ w_1 = (-1, 1)\,,~~~~\lambda_2 = e^{2 \phi}\,,~ w_2 = (1, 1)\,.
\end{equation}
The former eigenvector is the one of interest to us, as its eigenvalue vanishes in the limit $\phi \rightarrow \infty$. In other words, the field $B^- := - B^0 + B^1$ becomes strongly coupled in the limit $\phi \rightarrow \infty$, with gauge coupling $g_-^2 \sim e^{2 \phi}$. Meanwhile, $B^+ := B^0 + B^1$ becomes weakly coupled in this limit.

Let us assume that there is only one vector multiplet, $n_V =1$. We may then write the numbers $c_r^{11}$ simply as $c_r$. The last term in \eqref{action} may be written in the $B^\pm$ basis as
\begin{equation}
\frac{1}{2}  c_r B_{2}^r \wedge F_{2} \wedge F_{2} =  \frac{1}{2} F_{2} \wedge F_{2}  \wedge \left[ \frac{(c_0+c_1)}{2}  B^+_{2} + \frac{(-c_0 + c_1 )}{2} B^-_{2} \right]\,.
\end{equation}
For $c_0 \neq c_1$, this gives a nonzero Chern-Simons coupling, $B^- \wedge F \wedge F$, and consequently a non-trivial higher-group structure. The kinetic term for $F$ may then be written as
\begin{equation}
- \frac{1}{2} (c_0 \cosh (\phi)  - c_1 \sinh (\phi)) F_{2} \wedge \star F_{ 2}\,,
\label{Ff}
\end{equation}
where the relative minus sign comes from the relation $c_0 = - c^0$. For $c_0 \neq c_1$, the coefficient diverges as $e^\phi \sim g_A^{-2}$. Thus, we have the relation
\begin{equation}
g_- \sim g_A^{-2} \sim e^{ \phi}\,
\end{equation}
in the limit $\phi \rightarrow \infty$.

We next assume that the WGC is satisfied for strings charged magnetically under the two-forms $B^\pm$. We also assume that the charges are quantized such that there exists a string charged magnetically under $B^-$ but not under $B^+$. This quantization assumption is relatively mild: it will hold, for instance, provided the charges are quantized in either the $B^0$, $B^1$ basis or the $B^+$, $B^-$ basis. (In the former case, the string in question will have magnetic charges $q^0 = - q^1 \in \mathbb{Z}$, whereas in the later case the string will have magnetic charges $q^+ = 0$, $q^- \neq 0 \in \mathbb{Z}$.)

These assumptions together imply that in the limit $\phi \rightarrow \infty$, some string charged under $B^-$ should become tensionless, as the WGC implies that its tension must be bounded above as $T \lesssim g_-^{-1} \rightarrow 0$. 

Note that the quantization assumption will be satisfied provided the string charges are quantized in either the $B^0$, $B^1$ basis or the $B^\pm$ basis, so it 

Assuming a string charged magnetically under $B_-$ satisfies the WGC for strings, its string scale satisfies
\begin{equation}
M_{\textrm{string}} = \sqrt{2 \pi T } \lesssim g_-^{-1/2} \sim g_A  \,,
\label{relationeq}
\end{equation}
which means that the charged string excitations of the string satisfy the WGC, as expected in the presence of the $B^- \wedge F \wedge F$ Chern-Simons term.

On the other hand, if $c_0 = c_1$, then there is no Chern-Simons coupling $B^- \wedge F \wedge F$, and consequently no higher-group structure \eqref{mmixing} between $B^-$ and $F$. From \eqref{Ff}, the gauge coupling $g_A$ diverges as $g_A^2 \sim e^\phi$. Thus, the gauge field $A$ actually becomes strongly coupled in the limit $\phi \rightarrow 0$, whereas the string charged magnetically under $B^-$ becomes tensionless. As expected, the relation \eqref{relationeq} breaks down in the absence of the Chern-Simons coupling, and the excitations of the light string do not satisfy the WGC for the gauge field.  

A similar conclusion holds for $B^- \leftrightarrow B^+$ in the limit $\phi \rightarrow - \infty$.

\subsubsection{Two tensor multiplets}

To illustrate the robustness of the above results, we now consider the case of two tensor multiplets, $n_T = 2$, where there are two scalar fields, which parametrize the coset $SO(1,2)/SO(2)$. The equations \eqref{6dcond} can be solved with

\begin{align}
v_r&=\begin{pmatrix}\cosh(\phi_1),&\sinh(\phi_1)\sin(\phi_2),&\sinh(\phi_1)\cos(\phi_2) \end{pmatrix} \\
x^1_r &=\begin{pmatrix}\sinh(\phi_1),&\cosh(\phi_1)\sin(\phi_2),&\cosh(\phi_1)\cos(\phi_2) \end{pmatrix} \\
x^2_r &=\begin{pmatrix} 0 ,&\-\cos(\phi_2),&
-\sin(\phi_2) \end{pmatrix}.
\end{align}
Thus we have 

\begin{equation}
V = \begin{pmatrix} \cosh(\phi_1)&\sinh(\phi_1)\sin(\phi_2)&\sinh(\phi_1)\cos(\phi_2) \\ \sinh(\phi_1)&\cosh(\phi_1)\sin(\phi_2)&\cosh(\phi_1)\cos(\phi_2)\\ 0 &\-\cos(\phi_2)
&-\sin(\phi_2) \end{pmatrix}.
\end{equation}

This leads to a gauge kinetic matrix for the 2-form gauge fields of the form
 \begin{equation}
G_{rs} = \begin{pmatrix} 
\cosh(2 \phi_1)& \sin(\phi_2) \sinh(2 \phi_1)& \cos(\phi_2) \sinh(2 \phi_1)
 \\ 
 \sin(\phi_2) \sinh(2\phi_1)& 
\cosh^2(\phi_1) -\cos(2\phi_2) 
   \sinh^2(\phi_1) & 
  \sinh(\phi_1)^2 \sin(2\phi_2)
 \\ 
 \cos(\phi_2) \sinh(2 \phi_1)& \sinh^2(\phi_1) \sin(2 \phi_2)& 
  
  \cos(2 \phi_2) \sinh^2(\phi_1) + \cosh^2(\phi_1)
\end{pmatrix}
\end{equation}
with eigenvalues and eigenvectors:
\begin{align}
w_1 &=\begin{pmatrix} 1, & \sin(\phi_2),& \cos(\phi_2)  \end{pmatrix}  \text{ with } \lambda_1= e^{2 \phi_1} \nonumber  \\
w_2 &=\begin{pmatrix} -1 ,& \sin(\phi_2) ,& \cos(\phi_2)  \end{pmatrix}  \text{ with } \lambda_2= e^{-2 \phi_1}  \\
w_3&=\begin{pmatrix} 0,& -\cos(\phi_2),& \sin(\phi_2)  \end{pmatrix}  \text{ with } \lambda_3= 1  .\nonumber
\end{align}
Using these eigenvectors, we define 2-form gauge fields 
\begin{align}
B^+ &= B^0 + \sin( \phi_2) B^1 + \cos (\phi_2) B^2   \nonumber \\
 B^- &=  -B^0 + \sin( \phi_2) B^1 + \cos (\phi_2) B^2   \\
 B^{*} &=  - \cos(\phi_2) B^{1} + \sin ( \phi_2 ) B^2 \nonumber .
\end{align}
We are especially interested in $B^-$, which becomes strongly coupled in the limit $\phi_1 \rightarrow \infty$.

Meanwhile, the gauge kinetic term for the 1-form gauge field takes the form
\begin{align}
&-\frac{1}{2} ( \cosh(\phi_1)  c_0  -  \sinh(\phi_1)\sin(\phi_2) c_1  -  \sinh(\phi_1)\cos(\phi_2)  c_2 ) F_{2} \wedge \star F_{2} \nonumber  \\
&= -\frac{1}{2} \left( e^{\phi_1} (c_0 - c_1 \sin(\phi_2)  - c_2 \cos(\phi_2) )  +  e^{-\phi_1} (c_0 + c_1 \sin(\phi_2)  + c_2 \cos(\phi_2) )  \right) F_{2}  \wedge \star F_{2}
\label{kin6d}
\end{align}

The 1-form gauge field couples to the 2-forms via the Chern-Simons terms 
\begin{align}
 - \frac{1}{2}  F_{2} \wedge F_{2} \wedge &\Big[ \frac{(c_0+  c_1 \sin(\phi_2) +c_2 \cos(\phi_2)  )}{2}  B^+_{2}  + \frac{(-c_0 + c_1 \sin(\phi_2) +c_2 \cos(\phi_2) )}{2} B^-_{2} \nonumber \\
 &+( - c_1 \cos(\phi_2 + c_2 \sin(\phi_2) )  B^{*}_{2}  \Big]\,.
 \label{CS6d}
\end{align}

 For $c_0 \neq c_1 \sin(\phi_2) +c_2 \cos(\phi_2)$, we see from \eqref{CS6d} that $B^-$ couples to $F$ via a Chern-Simons coupling $B^- \wedge F \wedge F$, so there is a nontrivial higher-group structure. In addition, from \eqref{kin6d}, we see that the gauge coupling $g_A$ scales as $g_A^{-2} \sim e^{\phi_1}$. This implies a relation between the gauge coupling of $B^-$ and the gauge coupling $g_A$:
\begin{equation}
g_- \sim g_A^{-2} \sim e^{ \phi_1}\,
\end{equation}
in the limit $\phi_1 \rightarrow \infty$.

We now assume, as in the case of a single tensor multiplet above, that the WGC is satisfied for strings in the theory, and we assume that charges are quantized so that there exists a WGC-satisfying string charged magnetically only under $B^-$ (but not under $B^+$ or $B^*$). This latter assumption will be violated when $\tan(\phi_2)$ is irrational if charges are quantized in the original $B^0$, $B^1$, $B^2$ basis. However, it will be satisfied if charges are quantized in the $B^+$, $B^-$, $B^*$ basis. Furthermore, in F-theory compactifications to 6d, Lee, Lerche, and Weigand argued that any infinite distance, weak coupling limit of a 2-form gauge field which keeps gravity dynamical will lead to an asymptotically tensionless, weakly coupled heterotic string \cite{Lee:2019xtm}. Here, gravity indeed remains dynamical in the limit $\phi_1 \rightarrow \infty$, and the requirement that the asymptotically tensionless string is weakly coupled implies that it must be magnetically charged under $B^-$ only, since the magnetic duals of $B^+$ and $B^*$ are not weakly coupled in this limit. Thus, our assumption of a WGC-satisfying string charged magnetically under $B^-$ only appears to be satisfied in the context of 6d F-theory compactifications, which relies on a charge quantization condition that depends on additional input from compactification geometry rather than 6d supergravity alone.

Assuming there is such a WGC-satisfying string charged magnetically under $B^-$ only, its string scale satisfies
\begin{equation}
M_{\textrm{string}} = \sqrt{2 \pi T } \lesssim g_-^{-1/2} \sim g_A  \,,
\label{relationeqnT2}
\end{equation}
which means that the charged string excitations of the string satisfy the WGC, as expected from the presence of the $B^- \wedge F \wedge F$ Chern-Simons term.

On the other hand, for $c_0 = c_1 \sin(\phi_2) +c_2 \cos(\phi_2)$, we see from \eqref{CS6d} that $B^-$ does NOT couple to $F$ via a Chern-Simons coupling $B^- \wedge F \wedge F$, so $B^-$ and $F$ are not tied up into the higher-group structure of \eqref{mmixing}. Furthermore, from \eqref{kin6d}, we see that the gauge coupling $g_A$ scales as $g_A^{-2} \sim e^{- \phi_1} $, so $A$ actually becomes strongly coupled while the string charged magnetically under $B^-$ becomes  tensionless in the $\phi_1 \rightarrow 0$ limit. As expected, the relationship \eqref{relationeqnT2} does not hold in the absence of the Chern-Simons coupling, just as in the $n_T =1$ case above.

A similar conclusion holds for $B^- \leftrightarrow B^+$ in the limit $\phi_1 \rightarrow - \infty$.

\subsection{Strings in 7d supergravity}

We will now show how a similar story plays out in 7d supergravity. At first, we introduce and review the relevant aspects of 7 dimensional supergravity following the notation in \cite{Bergshoeff:1985mr}.

7d supergravity features one supergravity multiplet, and $n$ vector multiplets. The supergravity multiplet has one scalar field, three abelian vector fields, and one 2-form gauge field, whereas each vector multiplet has three scalar fields and one vector field. 
Thus, there are a total of $3n+1$ scalar fields: the dilaton $\sigma$ coming from the gravity multiplet, and $3n$ scalars $\phi^\alpha$ which come from the vector multiplets and parametrize the coset $SO(3, n) / (SO(3) \times SO(n))$. These may be described by the $(n+3) \times (n+3)$ matrix $L_I^A$ $(I,A=1,...,n+3)$ which satisfies the orthogonality condition
\begin{equation}
    L^T\eta L= \eta 
    \label{eq:L}
\end{equation}
where $\eta$ has signature $(-, -, -, +, ..., +)$.

It is useful to decompose $L_I^A$ into $n+3$ $Sp(1)$ matrices $(L^I)^i_j, i, j = 1,2$ and $n$ $SO(3, n)$ vectors $L^a_I$, $a = 1, ..., n$ \cite{Bergshoeff:1985mr}. The orthogonality condition in terms of these new matrices reads 
\begin{equation}
    -(L_I)^i_j (L_J)^j_i + L_I^a L^a_J= \eta_{IJ} \,.
\end{equation}
We also have matrices associated with the decomposition of the inverse matrix (defined as $L^{-1}=\eta L^T \eta $):
\begin{equation}
    (L^I)^i_j= - \eta^{IJ} [(L_J)^j_i]^* \,.
\end{equation}
We also define
\begin{equation}
    L^I_a= \eta^{IJ} L_J^a\,.
\end{equation}
 These satisfy the following relations:
 \begin{align}
     L^a_I (L^I)^i_j &=0 \\
L_I^a L^I_b&= \delta_b^a \\
(L_I)^{i}_{j} (L^I)^{k}_l&= \delta^i_l\delta^k_j  -\frac{1}{{2}}\delta^i_j\delta^k_l
\end{align}
We define 
\begin{equation}
(P_{a})^i_j  =  (P_{\mu, a})^i_j \rmd x^\mu = L_a^I \partial_\mu (L_{I})^i_j \rmd x^\mu
\label{7dP}
\end{equation}
as well as the modified field strength for the 2-form gauge field $B_{\mu \nu}$:
\begin{equation}
H_3  =  \rmd  B_2 - \frac{1}{3 \sqrt{2}}  \eta_{IJ} A_1^I \wedge F_2^J \,.
\end{equation}
The relevant part of the action can then be written as
\begin{equation}
  S= 
   \int 
   - \frac{1}{2} e^\sigma a_{IJ} F^I_{2} F_2^{ J} 
   -\frac{1}{2} e^{2\sigma} H_3 \wedge \star H_3 
   - \frac{5}{8} \rmd \sigma \wedge \star \rmd \sigma
   - \frac{1}{2} (P^a)^i_j \wedge \star (P_a)^j_i
   \label{7daction}
    \end{equation}
      where $a_{IJ}$ is defined as 
    \begin{equation}
        a_{IJ}= (L_I)^i_j (L_J)^j_i + L^a_I L_{J a}.
        \label{7da}
    \end{equation}

\subsubsection{No vector multiplets}

In the absence of vector multiplets, the action \eqref{7da} simplifies to 
\begin{equation}
S =\int  - \frac{1}{2} e^{2 \sigma } H_{3} \wedge \star H_{3} - \frac{1}{2} e^{\sigma} a_{IJ} F^I_{2} \wedge \star F_2^{J}  - \frac{5}{8} \rmd \sigma \wedge \star \rmd \sigma \,,
\end{equation}
where 
\begin{equation}
H_{3}  =  \rmd   B_{2} + \frac{1}{3 \sqrt{2}}  ( A_1^1 \wedge F_{2}^1 + A_1^2 \wedge F_{2}^2 + A_1^3 \wedge F_{2}^3  ) \,.
\label{Bmodification}
\end{equation}
Here, the gauge coupling $g_B$ for the 2-form $B_2$ scales with the gauge couplings $g_A := g_{A, I}$ for the 1-forms $A_1^I$ as 
\begin{equation}
g_B \sim g_A^2 \sim e^{2 \sigma}\,.
\end{equation}
This means that a string charged (electrically) under $B_2$ will have a string scale of the form
\begin{equation}
M_{\textrm{string}}  = \sqrt{2 \pi T}  \sim g_B^{1/2} \sim g_A  \, .
\label{7dnovector}
\end{equation} 
Due to the $A^I \wedge F^I$ term in \eqref{Bmodification}, the action features Chern-Simons couplings of the form $\star H_3 \wedge A^{I}_1 \wedge F^{I}_2$ for all three gauge fields, $I=1,2,3$. Dualizing $B_2$ to $C_3$, this leads to Chern-Simons couplings of the form $C_3 \wedge F_2^I \wedge F_2^I$.
By the usual anomaly inflow argument, this means that there are string excitations charged under each of the three gauge fields, which by \eqref{7dnovector} will satisfy the WGC, as expected.

\subsubsection{One vector multiplet}

Next, we consider a theory with a single vector multiplet, $n=1$. We parametrize the matrix $L_I^A$ from \eqref{eq:L} in terms of three scalar fields $\phi_i$ as follows:
\begin{align}
L_I^A&=
\begin{pmatrix}
 \cos(\phi_3) \cosh(\phi_1) \sin(\phi_2)& 
  \cos(\phi_2) \cos(\phi_3) \cosh(\phi_1)& 
  \cosh(\phi_1) \sin(\phi_3)& \sinh(\phi_1)&\\
  \cos(\phi_2)& -\sin(\phi_2) &0& 0 &\\
  \sin(\phi_2) \sin(\phi_3)& 
  \cos(\phi_2) \sin(\phi_3)& -\cos(\phi_3)&
   0&\\
   \cos(\phi_3) \sin(\phi_2) \sinh(\phi_1)&\cos(\phi_2) 
\cos(\phi_3) \sinh(\phi_1)&\sin(\phi_3) \sinh(\phi_1)& \cosh(\phi_1)
   \end{pmatrix}   
\end{align}
Here, $\phi_1$ can be viewed as a radial mode, whereas $\phi_2$ and $\phi_3$ are angular coordinates. For simplicity, we focus on a particular direction in field space, setting $\phi_2 = 0$, $\phi_3 = \pi/2$ and letting $\phi_1$ vary. We then have
\begin{align}
L_{1 j}^i = \frac{1}{\sqrt{2}} \begin{pmatrix}
   0&1\\
   1&0
    \end{pmatrix} &\,,~~~~
 L_{2 j}^i =  \frac{1}{\sqrt{2}}\begin{pmatrix}
   0&i\\
   -i&0
    \end{pmatrix} \nonumber \\
 L_{3 j}^i = \frac{1}{\sqrt{2}} \cosh(\phi_1) \begin{pmatrix}
  1&0\\
   0&-1
    \end{pmatrix} &\,,~~~~
 L_{4 j}^i =  \frac{1}{\sqrt{2}} \sinh(\phi_1) \begin{pmatrix}
  1&0\\
   0&-1
    \end{pmatrix}  \end{align}
    $$
L^1_I =\begin{pmatrix}
 0,& 
  0,& 
   \sinh(\phi_1),&
   \cosh(\phi_1) \end{pmatrix}  \, .
   $$
   The matrix $P_{\mu j}^i$ from \eqref{7dP} takes the form 
\begin{equation}
P_{\mu j}^i =
\frac{1}{\sqrt{2}} \begin{pmatrix}
    \partial_\mu \phi_1 & 0 &\\
  0& -\partial_\mu \phi_1 
\end{pmatrix}
\end{equation}
so the $\phi_1$ kinetic term is given by
\begin{equation}
    -\frac{1}{2} P_{ j}^i \wedge \star P_{ i}^j= -\frac{1}{2} \rmd \phi_1 \wedge \star \rmd \phi_1\,.
\end{equation}
   From \eqref{7da}, the gauge kinetic matrix takes the form
   \begin{equation}
a_{IJ}=  
  \begin{pmatrix}

 1 & 0 & 0 & 0 &\\
 
 0&1&  0&0&\\
 
 0 & 0 &\cosh(2\phi_1) & \sinh(2\phi_1)& \\
 
  0& 0 & \sinh(2\phi_1)& \cosh(2\phi_1)

  \end{pmatrix}
\end{equation}
This has eigenvalues/eigenvectors given by
\begin{align} 
\lambda_1 = 1 \,,  ~v_1&= \begin{pmatrix}
1,&
  0,& 
0,& 
0
  \end{pmatrix} \,,~~~~
   \lambda_2 = 1,  ~v_2= \begin{pmatrix}
  0,&
  1,&
  0,& 
  0
  \end{pmatrix} \\
   \lambda_3 = e^{-2 \phi_1},  
  ~ v_3 &= \begin{pmatrix}
  0,&
   0,&
   -1,&
 1
  \end{pmatrix} \,,~~~
  \lambda_4 =  e^{2 \phi_1},  
   ~v_4= \begin{pmatrix}
  0 ,& 0,&
 1,&
  1
  \end{pmatrix}
\end{align}
We therefore define the linear combinations:
\begin{equation}
A^\pm =  \frac{1}{\sqrt{2}} (A^3 \pm A^4)  \,.
\end{equation}
With this, the action \eqref{7daction} takes the form\footnote{The angular scalar fields $\phi_2$ and $\phi_3$ also have kinetic terms, but they are omitted here as they are irrelevant for our purposes.}
\begin{align}
 S&= 
  \int 
   - \frac{1}{2} e^\sigma  \left( F^1_2  \wedge \star F^1_2  + F^2_2 \wedge \star F^2_2  \right) - \frac{1}{2} e^\sigma  \left(e^{2 \phi_1} F^+_{2} \wedge \star F^{ +}_2  +e^{-2 \phi_1} F^-_{2} \wedge \star F^{ _2-}  \right) \nonumber \\
   &-\frac{1}2 e^{2\sigma} H_{3} \wedge \star H_3 
   - \frac{5}{8} \rmd \sigma \wedge \star \rmd \sigma
   - \frac{1}{2} \rmd \phi_1 \wedge \star \rmd \phi_1   \,,
    \end{align} 
    where
\begin{equation}
H_{3}  =   \rmd B_{2} + \frac{1}{3 \sqrt{2}}   \left( A^1 \wedge F_{2}^1  +  A^2_{1} \wedge F_{2}^2  + \frac{1}{2} (A^+_{1} \wedge F_{2}^- +  A^-_{1}  \wedge F_{2}^+ )  \right) \,.
\end{equation}
Upon dualizing the 2-form $B_2$ to a 3-form $C_3$, the modified field strength will lead to Chern-Simons terms of the form
\begin{equation}
C_3 \wedge ( F^1 \wedge F^1 + F^2  \wedge F^2 + F^- \wedge F^+ )\,.
\end{equation}
Note that the relevant Chern-Simons terms $C \wedge F \wedge F$ are present for $F=F^1$ and $F^2$, but they are absent for $F^\pm$: there is instead only a mixed Chern-Simons term $C \wedge F^+ \wedge F^-$.

Relatedly, the gauge coupling of $B_2$ scales with the gauge couplings $g_1$, $g_2$ for $A^1$ and $A^2$ as
\begin{equation}
g_B \sim g_1^2 = g_2^2 \sim e^{-\sigma}\,.
\label{eq:scaling8d}
\end{equation} 
As above, this means that string excitations of a string charged under $B_2$ carry charge under $A^1$ and $A^2$, and these charged strings excitations satisfy the WGC for the 1-form gauge fields provided the string satisfies the WGC for $B_2$. Here, similar to the 6d story above, we are assuming that the gauge charges are quantized so that there exist states charged under $A^0$ and $A^1$ but not $A^+$ or $A^-$.

On the other hand, the relation \eqref{eq:scaling8d} does not hold for the gauge couplings $g_\pm$, since these couplings scale with $\phi_1$ as
\begin{equation}
g_\pm^2 \sim e^{-\sigma \mp 2 \phi_1 }\, . 
\end{equation} 
Thus, as expected, the string excitations charged under $A^\pm$ need not satisfy the WGC in the limit $\sigma \rightarrow \infty$, since there is no $C \wedge F \wedge F$ coupling for these gauge fields.

It is worth noting, however, that there is a $C_3 \wedge F^+ \wedge F^-$ coupling, and correspondingly the string scale of a WGC-satisfying string charged magnetically under $C_3$ will scale as
\begin{equation}
M_{\textrm{string}}   \lesssim  g_B^{1/2} \sim (g_+ g_-)^{1/2}  \,. 
\label{mixedscaling}
\end{equation}
The presence of the mixed Chern-Simons term $C_3 \wedge F^+ \wedge F^-$ on the one hand and the geometric mean of $g_+$ and $g_-$ in this scaling relation on the other is quite tantalizing. Indeed, \cite{Heidenreich:2021yda} found this same connection between Kaluza-Klein and winding charges in 4d upon Kaluza-Klein reduction of 5d abelian gauge theory, and below we will see that the same scaling relation occurs in 8d and 9d supergravities with $C_{d-4} \wedge F^+ \wedge F^-$ Chern-Simons couplings. The upshot is that in the presence of the mixed Chern-Simons coupling $C \wedge F^+ \wedge F^-$, string excitations of the charged string will satisfy the WGC for either $A^+$ or $A^-$ (depending on the sign of $\phi_1$), but not both.

\subsection{Strings in 8d supergravity}

We now review relevant aspects of 8d supergravity, following \cite{Awada:1985ag}.

There are two types of multiplets in 8d supergravity: the supergravity multiplet, and the vector multiplet. The former features a graviton, a 4-form gauge field $C_4$, a dilaton $\sigma$, and two vector bosons $A_1^i$, $i=1, 2$. A vector multiplet features a vector boson $A_1$ and a pair of scalar fields $\phi^i$. In a theory with $n$ vector multiplets, the scalar fields $\phi^x$, $x=1,...,2n$ parametrize the coset space $SO(n, 2)/(SO(2) \times SO(n))$, and can be usefully represented by the matrices
\begin{equation}
L_I^i\,,~~L^{Ii}\,,~~~L_I^a\,,~~~L^{Ia}\,,
\end{equation}
where $I=1,...n+2$, $a=1,...,n$ and $i=1,2$. Let us define
\begin{equation}
L_I^{\pm} = \frac{1}{\sqrt{2}} ( L_I^1 \mathbf{1} + \Gamma_9 L_I^2)\,,~~~L^{I \pm} = \frac{1}{\sqrt{2}} ( L^{I1} \mathbf{1} + \Gamma_9 L^{I2})\,,
\end{equation}
with $\mathbf{1}$ the $8 \times 8$ identity matrix, and $\Gamma_9$ is  the Dirac gamma matrix satisfying $\Gamma_9^2 = -\mathbf{1}$, $\textrm{Tr}(\Gamma_9) = 0$. Then, the $L$'s satisfy
\begin{align}
\text{Tr} (L^{I \pm} L_I^\mp ) = 8\,,~~~L_I^a L^{Ib} = \delta^{ab}\nonumber \\
L^{I \pm} L_I^\pm = 0\,,~~~ L^{I \pm} L_I^a = 0\,,~~~L^{Ia} L_I^{\pm} = 0\,.
\label{8dLs}
\end{align}
Here, $I$ indices are raised and lowered using $\eta_{IJ}$, which we take to have signature $(-, -, ..., -, + , +)$.

The relevant part of the action takes the form
\begin{equation}
S = \int   -  \frac{1}{2} e^{-2 \sigma}  G_5 \wedge \star G_5 - \frac{1}{2} a_{IJ} e^{\sigma} F_2^I \wedge \star F_2^J  - \frac{1}{48} \rmd \sigma \wedge \star \rmd \sigma - \frac{1}{2} g_{xy} \rmd \phi^x \wedge \star \rmd \phi^y +\frac{1}{\sqrt{2}} c_{IJ} C_4 \wedge F_2^I \wedge F_2^J\,,
\end{equation}
where $G_5 = \rmd C_4$, and
\begin{equation}
a_{IJ} =  \frac{1}{8} \text{Tr} (L_{(I}^{ +} L_{J)}^- ) + L_{(I}^a L_{J)}^a\,,~~~c_{IJ} =  \frac{1}{8} \text{Tr} (L_{(I}^{ +} L_{J)}^- ) - L_{(I}^a L_{J)}^a\, .
\end{equation}

\subsubsection{No vector multiplets}

In the absence of vector multiplets, the relevant part of the action takes the simple form
\begin{align}
S&= \int  -  \frac{1}{2} e^{-2 \sigma}  G_5 \wedge \star G_5 - \frac{1}{2} e^{\sigma} (F_2^1 \wedge \star F_2^1+ F_2^2 \wedge \star F_2^2)  - \frac{1}{48} \rmd \sigma \wedge \star \rmd \sigma  \nonumber \\
&+ \frac{1}{ \sqrt{2}} C_4 \wedge ( F_2^1 \wedge F_2^1 + F_2^2 \wedge F_2^2)\,.
\end{align}
In the limit $\sigma \rightarrow \infty$, the gauge couplings $g_1$, $g_2$ for the gauge fields $A^1$, $A^2$ scale as $e^{- \sigma/2}$, whereas the gauge coupling $g_B$ for the 2-form gauge field $B_2$ dual to $C_4$ scales as
\begin{equation}
g_B \sim g_1^2 \sim g_2^2 \sim e^{-\sigma}\,.
\end{equation}
Due to the presence of the Chern-Simons couplings $C_4 \wedge F^1 \wedge F^1$ and $C_4 \wedge F^2 \wedge F^2$, the excitations of a string charged under $B_2$ will also be charged under the 1-form gauge fields. If the string satisfies the WGC for $B_2$, then its excitations will have mass of order the string scale,
\begin{equation}
M_{\textrm{string}} = \sqrt{2 \pi T} \sim g_B^{1/2} \sim g_1 \sim g_2 \,,
\end{equation}
which means that these string states will satisfy the WGC for the 1-form gauge fields. Once again, we see that WGC mixing occurs in the presence of the Chern-Simons terms.

\subsubsection{One vector multiplet}

Defining $L_I^{a=1} := L_I^0$, the equations \eqref{8dLs} are solved by
\begin{align}
L_I^0&=\begin{pmatrix}\cosh(\phi_1),&\sinh(\phi_1)\sin(\phi_2),&\sinh(\phi_1)\cos(\phi_2) \end{pmatrix} \nonumber \\
L_I^1&= \sqrt{2} \begin{pmatrix}\sinh(\phi_1),&\cosh(\phi_1)\sin(\phi_2),&\cosh(\phi_1)\cos(\phi_2) \end{pmatrix}  \\
L_I^2&=\begin{pmatrix}0,& \cos(\phi_2),& - \sin(\phi_2) \end{pmatrix} \nonumber 
\end{align}
These give
 \begin{equation}
a_{IJ} = \begin{pmatrix} 
\cosh(2 \phi_1)& \sin(\phi_2) \sinh(2 \phi_1)& \cos(\phi_2) \sinh(2 \phi_1)
 \\ 
 \sin(\phi_2) \sinh(2\phi_1)& 
\cosh(2 \phi_1) \sin^2(\phi_2)+ \cos^2(\phi_2) 
 & 
  \sinh(\phi_1)^2 \sin(2\phi_2)
 \\ 
 \cos(\phi_2) \sinh(2 \phi_1)& \sinh^2(\phi_1) \sin(2 \phi_2)& 
  
  \cos^2( \phi_2) \cosh( 2 \phi_1) + \sin^2(\phi_2)
\end{pmatrix}\,.
\end{equation}
The associated eigenvalues and eigenvectors are 
\begin{align}
w_1 &=\begin{pmatrix} 1, & \sin(\phi_2),& \cos(\phi_2)  \end{pmatrix}  \text{ with } \lambda_1= e^{2 \phi_1} \nonumber  \\
w_2 &=\begin{pmatrix} -1 ,& \sin(\phi_2) ,& \cos(\phi_2)  \end{pmatrix}  \text{ with } \lambda_2= e^{-2 \phi_1}  \\
w_3&=\begin{pmatrix} 0,& -\cos(\phi_2),& \sin(\phi_2)  \end{pmatrix}  \text{ with } \lambda_3= 1  .\nonumber
\end{align}
Using these eigenvectors, we define gauge fields 
\begin{align}
A^+ &= A^0 + \sin( \phi_2) A^1 + \cos (\phi_2) A^2   \nonumber \\
 A^- &=  -A^0 + \sin( \phi_2) A^1 + \cos (\phi_2) A^2   \\
 A^{*} &=  - \cos(\phi_2) A^{1} + \sin ( \phi_2 ) A^2 \nonumber .
\end{align}
In terms of these gauge fields, there is no kinetic mixing, and their associated gauge couplings scale as 
\begin{equation}
g_+^{2} \sim e^{-\sigma - 2 \phi_1}\,,~~~g_-^{2} \sim e^{-\sigma + 2 \phi_1}\,,~~~g_*^2 \sim e^{- \sigma}\,.
\end{equation}
The Chern-Simons couplings are determined by the matrix $C_{IJ} = \eta_{IJ} = \text{diag}(-1, 1, 1)$ to be
\begin{equation}
\frac{1}{ \sqrt{2}} C_4 \wedge \left( F^{*} \wedge F^{*}  + F^+ \wedge F^- \right)\,.
\end{equation}
The gauge coupling of the 2-form $B_2$ dual to $C_4$ then obeys
\begin{equation}
g_B  \sim g_*^2 
\label{eq:8dscale}
\end{equation}
as in pure supergravity, and the $C_4 \wedge F_2^{*} \wedge F_2^{*}$ Chern-Simons term implies that the excitations of a superextremal string will satisfy the WGC for $A_1^{*}$, assuming that charges are quantized so that there exist states charged under $A^*$ but not under $A^+$ or $A^-$.

On the other hand, there is no $C_4 \wedge F_2^\pm \wedge F_2^\pm$ Chern-Simons coupling, and consequently the string excitations of a string charged under $B_2$ need not satisfy the WGC for $A_1^\pm$. Relatedly, as expected, the scaling \eqref{eq:8dscale} does not apply with $g_*$ replaced by $g_\pm$, since these gauge couplings scale not only with $\sigma$, but also with $\phi_1$. There is a $C_4 \wedge F_2^+ \wedge F_2^-$ coupling, and correspondingly there is a relation between the gauge couplings $g_B$, $g_\pm$ of the form $
g_B  \sim  g_+ g_-  $, as we saw in 7d supergravity in \eqref{mixedscaling}. This ensures that the excitations of a string which satisfies the WGC for $B_2$ will generically satisfy the WGC for exactly one of the gauge fields $A^+$ and $A^-$, depending on the sign of $\phi_1$.

\subsection{Strings in 9d supergravity}

9d supergravity is very similar to the cases we have already seen above. The only multiplets are the supergravity multiplet and the vector multiplet. The former features a graviton, a 2-form gauge field, a 1-form gauge field, and a scalar. The latter contains a vector field and a scalar field. Together, the scalar fields of a  supergravity multiplet and $n$ vector multiplets parametrize the coset space $SO(1, n)/SO(n)$.

In pure supergravity, the action for the gauge fields takes the form
\begin{equation}
S = \int - \frac{1}{2} e^{2 \sigma} H_3 \wedge \star H_3 - \frac{1}{2} e^\sigma F_2 \wedge \star F_2 \,,
\end{equation}
where $H_3 =  \rmd B_2 - A_1 \wedge F_2$. The modified 3-form field strength $H_3$ implies a Chern-Simons coupling $C_5 \wedge F_2 \wedge F_2$, where $C_5$ is the magnetic dual of $B_2$, so excitations of a string charged under $B_2$ will carry charge under $A_1$. From the action, we have the expected behavior
\begin{equation}
g_B \sim g_A^{2} \sim e^{-2 \sigma}\,,
\label{eq:9dscaling}
\end{equation} 
so as $\sigma \rightarrow \infty$, string excitations of a string which satisfies the WGC for the 2-form $B_2$ will have mass of order
\begin{equation}
M_{\textrm{string}} = \sqrt{2 \pi T} \lesssim g_B^{1/2} \sim g_A\,,
\end{equation}
so these excitations will satisfy the WGC for $A_1$.

With one vector multiplet, on the other hand, the action takes the form
\begin{equation}
S  = \int - \frac{1}{2} e^{2 \sigma} H_3 \wedge \star H_3 - \frac{1}{2} e^{\sigma+ 2 \phi} F_2^+ \wedge \star F_2^+   - \frac{1}{2} e^{\sigma- 2 \phi} F_2^- \wedge \star F_2^- \,,
\end{equation}
where $H_3 = \rmd B_2 - A_1^+ \wedge F_2^-$. The gauge couplings for $B_2$, $A_1^\pm$ scale as
\begin{equation}
g_B^2 \sim e^{-2 \sigma}\,,~~~g_+^2 \sim e^{-\sigma - 2 \phi }\,,~~~g_-^2 \sim e^{-\sigma + 2 \phi}\,,.
\end{equation}
The relationship \eqref{eq:9dscaling} does not hold for either $g_+$ or $g_-$, as each of these gauge couplings scale not only with $\sigma$, but also with $\phi$. Relatedly, there is no $C_5 \wedge F_2 \wedge F_2$ coupling for these gauge fields, but there is a there is a $C_5 \wedge F_2^+ \wedge F_2^-$ coupling, and correspondingly there is a relation between the gauge couplings $g_B, g_\pm$ of the form $g_B  \sim  g_+ g_- $, as expected.

\subsection{Strings in 10d supergravity}

Finally, we turn our attention to 10d supergravity. There are three types of supergravity to consider: Type IIA, Type IIB, Type I. Type IIB supergravity does not have a 1-form gauge field, so it is not relevant for our purposes. Thus, we restrict our analysis to Type IIA and Type I supergravity.

The actions for these theories can be found in \cite{Polchinski:1998rr}. After converting to Einstein frame, the relevant terms in the respective actions take the form
\begin{align}
S_{\textrm{IIA}} &=  \frac{1}{2} \left( \int  - \frac{1}{2} \rmd \sigma \wedge \star \rmd \sigma - \frac{1}{2} e^{\sigma} H_3 \wedge \star H_3  - \frac{1}{2}  e^{-3 \sigma/2} F_2 \wedge \star F_2   \right)  \\
S_{\textrm{I}} &= \frac{1}{2} \left( \int  - \frac{1}{2} \rmd \sigma \wedge \star \rmd \sigma - \frac{1}{2} e^{\sigma} H_3 \wedge \star H_3  - \frac{1}{g_A^2} e^{ \sigma/2}   \textrm{Tr} (F_2 \wedge \star F_2)  \right) \,. 
\end{align}
Here, the gauge group associated with the connection $A_1$ is necessarily $U(1)$ for Type IIA supergravity, but for Type I supergravity it is an unspecified nonabelian Lie group.

Crucially, the definitions of the 3-form $H_3$ differ between Type IIA and Type I supergravity. In Type IIA, we have simply $H_3 = \rmd B_2$, whereas in Type I, we have
\begin{equation}
\textrm{Type I:}~~H_3 = \rmd B_2  - \frac{1}{g_A^2} \textrm{Tr} ( A_1 \wedge \rmd A_1 - \frac{2i}{3} A_1 \wedge A_1 \wedge A_1 ) \,.  
\end{equation}
Upon dualizing $B_2$ to a 6-form $C_6$, this leads to a Chern-Simons coupling of the form $C_{6} \wedge \textrm{Tr} (F_2 \wedge F_2)$. This coupling is absent in Type IIA supergravity.

We are interested in the scaling of the gauge coupling $g_B$ with $g_A$. From the form of the respective actions, we have
\begin{equation}
\textrm{Type IIA:}~~ g_B \sim g_A^{-2/3} \sim e^{-\sigma/2}\,,~~~~~~
\textrm{Type I:}~~ g_B \sim g_A^2 \sim e^{-\sigma/2}\,.
\end{equation}
As $\sigma \rightarrow \infty$ in Type I supergravity, we see the expected scaling $g_B \sim g_A^2$, hence the string excitations of a string which satisfies the WGC for the 2-form $B_2$ will satisfy the WGC for the 1-form $A_1$, by the usual argument.

In contrast, as $\sigma \rightarrow \infty$ in Type IIA supergravity, the 1-form gauge field actually becomes strongly coupled while the 2-form $B_2$ becomes weakly coupled. This is unsurprising, however, since there is no $C_6 \wedge F_2 \wedge F_2$ Chern-Simons coupling, hence no WGC mixing is expected.

\section{BF Theory}\label{sec:BF}

In this section, we consider theories with two-term Chern-Simons couplings, also known as BF theories. Our starting point is the action
\begin{equation}
S = \int_X -\frac{1}{2 g_A^2} F_{p+1} \wedge \star F_{p+1} - \frac{1}{2 g_B^2} H_{d-p} \wedge \star H_{d-p} +\frac{ K }{2 \pi} B_{d-p-1} \wedge F_{p+1} \,,
\label{BF}
\end{equation}
where $F_{p+1} = \rmd A_{p}$, $H_{d-p} = \rmd B_{d-p-1}$. Without the BF coupling (i.e., for $K=0$), the system has four higher form symmetries, with conserved currents
\begin{align}
\frac{1}{2 g_A^2} {\star F_{p+1}} \,,~~~ \frac{1}{2 g_B^2} {\star H_{d-p}} \,,~~~\frac{1}{2 \pi} F_{p+1}\,,~~~\frac{1}{2 \pi} H_{d-p}\,.
\end{align}
The first two of these generate electric symmetries of $A_{p+1}$ and $B_{d-p-1}$, respectively, whereas the latter two generate magnetic symmetries. In the presence of the BF coupling (i.e., for $K \neq 0$), the magnetic symmetries are gauged, and the electric symmetries are each broken to a $\mathbb{Z}_K$ subgroup. The action can then be written as 
\begin{align}
S &= \int_X \Big[- \frac{1}{2 g_A^2}  (F_{p+1} - a_{p+1}^{(e)}) \wedge \star (F_{p+1} - a_{p+1}^{(e)}) - \frac{1}{2 g_B^2} (H_{d-p} -   b_{d-p}^{(e)}) \wedge \star (H_{d-p} -   b_{d-p}^{(e)})   \Big] \nonumber \\ 
& + \frac{ K }{ 2 \pi}  \int_Y(H_{d-p}   -   b_{d-p}^{(e)})  \wedge (F_{p+1}  - a_{p+1}^{(e)}) \,,
\end{align}
Here, $Y$ is a $(d+1)$-manifold whose boundary is $X$, and the background gauge fields satisfy
\begin{equation}
K a_{p+1}^{(e)} = \rmd \lambda_{p}^{(a)}  \,,~~~~ K b_{d-p}^{(e)} = \rmd \lambda_{d-p-1}^{(b)}\,,
\label{Kgaugeeq}
\end{equation}
This means that the background fields are flat gauge fields whose holonomies are $K$th roots of unity.
Mathematically, we may describe the gauge invariant information of these background gauge fields in terms of cohomology classes
\begin{equation}
[\frac{K}{2 \pi} a_{p+1}^{(e)} ] \in H^{p+1}(X, \mathbb{Z}_K)\,,~~~[\frac{K}{2 \pi} b_{d-p}^{(e)} ] \in H^{d-p}(X, \mathbb{Z}_K)\,.
\end{equation}
The electric background gauge fields transform simply as
\begin{equation}
a_{p+1}^{(e)} \rightarrow  a_{p+1}^{(e)} + \rmd \Lambda_p^{(a, e)}\,,~~~b_{d-p}^{(e)} \rightarrow  b_{d-p}^{(e)} + \rmd \Lambda_{d-p-1}^{(b,e)}\,,
\end{equation}
which means that the gauge parameters $\lambda_{p}^{(a)}$, $\lambda_{d-p-1}^{(b)} $ transform as
\begin{equation}
\lambda_{p}^{(a)} \rightarrow  \lambda_{p}^{(a)} + K \Lambda_p^{(a, e)}\,,~~~\lambda_{d-p-1}^{(b)} \rightarrow  \lambda_{d-p-1}^{(b)} + K \Lambda_{d-p-1}^{(b,e)}\,.
\end{equation}
Since the magnetic symmetries are gauged, there is no possibility of a higher-group structure.

Things become slightly more interesting if we couple the system to dynamical currents $j_{d-p}^a$, $j_{p+1}^b$, which couple to $A_p$ and $B_{d-p-1}$, respectively. In this case, the gauged currents are no longer $\frac{1}{2 \pi} F_{p+1}$ and $\frac{1}{2 \pi} H_{d-p}$, but rather the linear combinations
\begin{equation}
\frac{K}{2 \pi} F_{p+1} + j_{p+1}^b\,,~~~\frac{K}{2 \pi} H_{d-p} + j_{d-p}^a\,.
\end{equation}
$\frac{1}{2 \pi} F_{p+1}$ and $\frac{1}{2 \pi} H_{d-p}$ remain as valid symmetries of the theory, though they will be broken in the presence of magnetic monopoles for $A_p$ and $B_{d-p-1}$, respectively.

We may couple this system to background gauge fields as
\begin{align}
S &= \int_X \Big[ -\frac{1}{2 g_A^2} (F_{p+1} - a_{p+1}^{(e)}) \wedge \star (F_{p+1} - a_{p+1}^{(e)}) - \frac{1}{2 g_B^2} (H_{d-p} -   b_{d-p}^{(e)}) \wedge \star (H_{d-p} -   b_{d-p}^{(e)})  \nonumber \\
&+ \frac{1}{2 \pi} A_{p} \wedge f_{d-p}^{(m)}   +  \frac{1}{2 \pi} B_{d-p-1} \wedge h_{p+1}^{(m)}  \Big]  + \frac{K }{ 2 \pi}  \int_Y (H_{d-p}  + \frac{2 \pi }{K} j_{d-p}^a -   b_{d-p}^{(e)})  \wedge (F_{p+1} +\frac{2 \pi }{K} j_{p+1}^b - a_{p+1}^{(e)}) \,.
\end{align}
Once again, $Y$ is a $(d+1)$-manifold whose boundary is $X$.

The electric background gauge fields transform simply as
\begin{equation}
a_{p+1}^{(e)} \rightarrow  a_{p+1}^{(e)} + \rmd \Lambda_p^{(a, e)}\,,~~~b_{d-p}^{(e)} \rightarrow  b_{d-p}^{(e)} + \rmd \Lambda_{d-p-1}^{(b,e)}\,,
\end{equation}
but the gauge transformations for the magnetic background fields would seem to be mixed up into a higher-group-like\footnote{We use the term ``higher-group-like" because the form degrees of $\Lambda^{b ,e}_{d-p-1}$ and $\Lambda^{a,e}_{p}$ appearing on the right-hand side are too large to fit the definition of a higher-group \eqref{highergroup}.} structure:
\begin{align}
a_{d-p-1}^{(m)} &\rightarrow  a_{d-p-1}^{(m)} + \rmd \Lambda_{d-p-2}^{(a, m)} - K \Lambda_{d-p-1}^{(b,e)}   \\
b_{p}^{(m)} &\rightarrow  b_{p}^{(m)} + \rmd \Lambda_{p-1}^{(b,m)} - K \Lambda_{p}^{(a,e)} \,,
\label{BFmmixing}
\end{align}
with gauge-invariant field strengths given by
\begin{align}
f_{d-p}^{(m)} &= \rmd a_{d-p-1}^{(m)} + K b_{d-p}^{(e)}  \\
h_{p+1}^{(m)} &= \rmd b_{p}^{(m)} + K a_{p+1}^{(e)}   \, . \label{eqBFh}
\end{align}
However, we see from \eqref{Kgaugeeq} that the terms $  K b_{d-p}^{(e)}$ and $ K a_{p+1}^{(e)} $ on the right-hand side of these equations are pure gauge. In other words, they contain no nontrivial gauge-invariant information, and they can be set trivial by a gauge transformation. There is no higher-group structure to speak of, even in the presence of the dynamical currents $j_{d-p}^a$, $j_{p+1}^b$, and the electric $\mathbb{Z}_K$ symmetries may exist even when the magnetic $U(1)$ symmetries are broken.

\subsection{Abelian Higgs model}

Despite the absence of a higher-group structure, there remains an interesting interplay between weak gravity conjectures for different gauge fields in the presence of the BF coupling. Consider a particular ultraviolet completion of BF theory in four dimensions: namely, the abelian Higgs model with a complex scalar of charge $K$. The action takes the form
\begin{equation}
S = \int - \frac{1}{2 e^2} F \wedge \star F - \rmd_A \Phi^\dagger \wedge \star \rmd_A \Phi - V(\Phi)\,,
\end{equation}
with $\rmd_A$ the covariant derivative,
\begin{equation}
\rmd_A = \rmd + i K A_1
\end{equation}
and
\begin{equation}
V(\Phi) = \frac{1}{2} \lambda \left( |\Phi|^2 - f^2 \right)^2\,.
\end{equation}
We assume that we are working at sufficiently low energies, so that we may integrate out the radial mode, and we are left with a (non-fundamental) axion $\theta$, of decay constant $f$, coupled to the gauge field via a Stueckelberg coupling. Dualizing the axion, we find a 2-form $B_2$, which couples to $A_1$ via a BF coupling at level $K$, as in \eqref{BF}.

However, if we want to discuss the WGC, we cannot work at energies that are too low. At energies below the mass of the photon and 2-form gauge field,
\begin{equation}
m_A \sim m_B \sim K e f \,,
\end{equation}
the $U(1)$ gauge symmetries for $A_1$ and $B_2$ are broken to $\mathbb{Z}_K$ subgroups. For $K=1$, the gauge symmetries are thus completely broken. There is at present no well-justified analog of the WGC for discrete gauge groups, and there is certainly no WGC statement for trivial gauge groups, so the WGC is vacuous at very low energies.

However, it is generally believed that the WGC should be satisfied for a massive $p$-form gauge field at energies well above the mass of the gauge field. This may be related to the fact that a $U(1)$ electric $(p+1)$-form symmetry is restored at such energies, and the WGC may be related to the breaking of higher-form global symmetries \cite{CordovaOhmoriRudelius}.

As a result, in order for the WGC to provide a nontrivial bound on strings charged under $B_2$, we must have a parametric separation of energy scales
\begin{equation}
K e f \ll \sqrt{\lambda} f\,.
\label{separation}
\end{equation}
If this inequality is not satisfied, then there is no energy scale at which $B_2$ behaves as a $U(1)$ 2-form gauge field, and there is no energy scale at which an approximate $U(1)$ electric 2-form symmetry exists, so the WGC for strings is satisfied trivially.

If the scale separation in \eqref{separation} is obeyed, however, then the 2-form WGC requires the existence of a charged string whose tension satisfies
\begin{equation}
T \lesssim f \Mpl\,.
\label{WGC1}
\end{equation}
The tension of such a string can be estimated as $T \sim f^2$ \cite{Dolan:2017vmn, Hebecker:2017wsu}, so the 2-form WGC bound \eqref{WGC1} becomes
\begin{equation}
f \lesssim \Mpl \,.
\label{2formwgc}
\end{equation}
This bound will be satisfied provided the Higgs field satisfies the WGC for the 1-form $A_1$:
\begin{equation}
\sqrt{\lambda} f \sim m_\Phi \lesssim K e \Mpl\,,
\end{equation}
since together with \eqref{separation} this implies the 2-form WGC bound \eqref{2formwgc}. Thus, we see that a sort of WGC mixing persists in the presence of the two-term Chern-Simons term, $B \wedge F$.

As in \S\ref{Strings} above, we are making some assumptions here that go beyond the statements of the WGCs themselves. We are assuming that we have a Stueckelberg description, which completes at high energies to an abelian Higgs model. We are assuming that the 1-form and 2-form WGCs can be applied at energy scales well above the mass of $A_1$ and $B_2$, and we are assuming that an (intermediate) energy range exists over which $B_2$ behaves like a $U(1)$ gauge field. We are assuming that the Higgs field itself satisfies the WGC.

It is worth noting that the energy scale of the charged string $M_{\textrm{string}} = \sqrt{ 2 \pi T} \sim f$ here may be well above the energy scale $\sqrt{\lambda } f$ at which the 2-form symmetry itself breaks down. Perhaps this means that the WGC need not apply to strings charged under the gauge field $B_2$. A better understanding of the domain of validity of the WGC is needed, and we leave this to future work.

\subsection{Axion monodromy}

Finally, we consider axion monodromy, which may be described in terms of a 4-form flux $F_4 = \rmd C_3$ \cite{Kaloper:2008fb}:
\begin{equation}
S =  \int - \frac{f_\theta^2}{2}  \rmd \theta \wedge \star \rmd \theta - \frac{1}{2 g_C^2} G_4 \wedge \star G_4 + \frac{ 1}{ 2 \pi } \theta G_4  \,,
\end{equation}
where $G_4 = \rmd C_3$, and $\theta$ is periodic under $\theta \rightarrow \theta + 2 \pi$.
This gives rise to a multi-branched potential for the axion of the form
\begin{equation}
V(\phi) = \frac{g_C^2}{2} (1 + \frac{\theta}{2 \pi})^2 + \Lambda^4 \cos( \theta )\,,
\label{V}
\end{equation}
where $\Lambda^4 = e^{-S} \Lambda_{\text{UV}}^4$, where $S$ is the instanton action and $ \Lambda_{\text{UV}}$ is some UV scale, such as the Planck scale or the string scale.

As expected for a BF theory, the gauge fields $\theta$ and $C_3$ acquire masses from the $\theta G_4$ coupling,
\begin{equation}
m_\theta \sim m_C \sim g_C / f_\theta\,.
\end{equation}
Below this energy scale, then, the gauge symmetries of $\theta$ and $C_3$ are spontaneously broken, and their associated 0-form and 3-form electric symmetries are explictly broken. Any application of the WGCs for these gauge fields must therefore involve energy scales which are large compared to $g_C/f_\theta$: the WGC is vacuous at energies below this.

The WGC for the axion implies
\begin{equation}
f_\theta S < \Mpl \,.
\end{equation}
Meanwhile, the WGC for $C_3$ implies the existence of a domain wall with tension $T$ satisfying
\begin{equation}
T \lesssim g_C \Mpl\,,
\end{equation}
which interpolates between the different branches of the potential. The tension of this domain wall is sensitive to ultraviolet physics. In order to uncover any relationships between these two WGCs, therefore, we seemingly require input from the UV completion of this system, just as how the previous subsection relied on input from the UV completion of BF theory, namely, the abelian Higgs model. At present, however, we are not aware of any UV completion of $\phi G_4$ theory into a four-dimensional EFT. It would be worthwhile to explore this possibility, but we leave this for future work.

\section{Kinetic Alignment}\label{ALIGNMENT}

In this section, we consider gauge kinetic mixing of two $p$-form gauge fields, which for $p=0$ reduces to the decay constant alignment scenario of Kim, Nilles, and Peloso (KNP) \cite{kim:2004rp}. This scenario is perhaps already familiar to the reader, but here we describe it in a way that makes its connection to higher-group global symmetries more explicit.

We begin with the action for two $p$-form gauge fields, $A_p$ and $B_p$, with associated field strengths $F_{p+1} = \rmd A_p$ and $H_{p+1} = \rmd B_p$ and gauge couplings $g_A$, $g_B$:
\begin{equation}
S = \int -\frac{1}{2 g_A^2} F_{p+1} \wedge \star F_{p+1} - \frac{1}{2 g_B^2} H_{p+1} \wedge \star H_{p+1}\,.
\label{unmixed}
\end{equation}
The gauge fields have associated transformations laws
\begin{equation}
A_p \rightarrow A_p + \rmd \Lambda_{p-1}^{A}\,,~~~~B_p \rightarrow B_p + \rmd \Lambda_{p-1}^{B}\,.
\end{equation}
We further introduce background gauge fields $a_{p+1}$ and $b_{p+1}$ for the (electric) $p$-form symmetries associated with $A$ and $B$, respectively. The action with these gauge fields included is given by
\begin{equation}
S = \int -\frac{1}{2 g_A^2} (F_{p+1} - a_{p+1}) \wedge \star (F_{p+1} - a_{p+1}) - \frac{1}{2 g_B^2} (H_{p+1}-b_{p+1}) \wedge \star (H_{p+1} - b_{p+1})\,.
\end{equation}
The background gauge fields have associated gauge transformations
\begin{equation}
a_{p+1} \rightarrow a_{p+1} + \rmd \Lambda_{p}^{a}\,,~~~~b_{p+1} \rightarrow b_{p+1} + \rmd \Lambda_{p}^{b}\,.
\end{equation}

Now, we make a field redefinition, moving into a basis with kinetic mixing. More specifically, we define
\begin{equation}
A_{p}' = A_p\,,~~~~B_p' = B_p - \vartheta A_p\,.
\end{equation}
so that
$
F_{p+1}' = F_{p+1}$, $H_{p+1}' = H_{p+1} - \vartheta F_{p+1}
$.
In this new basis, the action \eqref{unmixed} becomes
\begin{equation}
S = \int - \frac{1}{2 (g'_A)^2} F'_{p+1} \wedge \star F'_{p+1} - \frac{1}{2 g_B^2} H'_{p+1} \wedge \star H'_{p+1} + \frac{\vartheta}{ 2g_B^2  } ( F'_{p+1} \wedge \star H'_{p+1} +  H'_{p+1}  \wedge \star F'_{p+1}  )\,,
\end{equation}
where
\begin{equation}
\frac{1}{(g_A')^2} = \frac{1}{g_A^2} + \frac{\vartheta^2}{g_B^2}\,.
\end{equation}
In the $A'$, $B'$ basis, the gauge kinetic matrix is given by
\begin{equation}
\tau_{ij} = \left( \begin{array}{cc}
\frac{1}{(g_A')^2} & -\frac{\vartheta}{g_B^2} \\
-\frac{\vartheta}{g_B^2}  &   \frac{1}{ g_B^2 } 
\end{array} \right) \,.
\end{equation}
If $\vartheta$ is taken to be very large, then $1/(g_A')^2$ may be much larger than $1/g_A^2$ and $1/g_B^2$, and one eigenvalue of $\tau_{ij}$ can be parametrically large even though the product of eigenvalues remains fixed, as $\det(\tau_{ij}) = (g_A g_B)^{-2}$.
In the case of $p = 0$ in 4d, where the gauge fields in question are axions, this behavior is the hallmark of decay constant alignment, also known as KNP alignment \cite{kim:2004rp}. In this scenario, the eigenvector with the larger eigenvalue serves as the inflaton field, ideally realizing a  model of large-field inflation with a super-Planckian traversal.

This kinetic mixing has consequences for the background gauge fields as well. We may write the action with background gauge fields included as
\begin{align}
S =& -\int\Big[ \frac{1}{2 (g_A')^2} (F'_{p+1} - a'_{p+1}) \wedge \star (F'_{p+1} - a'_{p+1}) \nonumber \\
&+ \frac{1}{2 g_B^2} (H'_{p+1} + \vartheta F'_{p+1} -b'_{p+1}) \wedge \star (H'_{p+1} + \vartheta F'_{p+1} - b'_{p+1}) \Big] \,,
\end{align}
where
\begin{equation}
a'_{p+1}= \frac{(g_A')^2}{ g_A^2}  a_{p+1}  + \vartheta \frac{(g_A')^2}{g_B^2} b_{p+1}\,,~~~~b'_{p+1} = b_{p+1}\,.
\end{equation}
These background gauge fields then transform under background gauge transformations as
\begin{equation}
a'_{p+1}  \rightarrow a'_{p+1} + \frac{(g_A')^2}{ g_A^2}  \rmd \Lambda_p^a +  \vartheta \frac{(g_A')^2}{g_B^2} \rmd \Lambda_p^b \, ,~~~~ b'_{p+1}  \rightarrow b'_{p+1} + \rmd \Lambda_p^b \, .
\end{equation}
Here, we see that the kinetic mixing of the gauge fields implies mixing of the gauge transformations $\Lambda_p^a$, $\Lambda_p^b$ for the background gauge field $a_{p+1}'$. This is reminiscent of the higher-group structure of \eqref{highergroup}.

The astute reader will note that this higher-group-like structure is a bit of an illusion, since we may also define modified gauge parameters
\begin{equation}
\Lambda_p^{a \, \prime}  = \frac{(g_A')^2}{ g_A^2}  \Lambda_p^a +  \vartheta \frac{(g_A')^2}{g_B^2} \Lambda_p^b\,,~~~  \Lambda_p^{b \, \prime} = \Lambda_p^b \,.
\label{lambdaprime}
\end{equation}
However, this redefinition comes at a cost: the periodicities of the modified gauge parameters will also be modified. Let us focus on the case $p=0$, since this is the case of interest for the purposes of axion inflation. Here, just as the axions $A_0$, $B_0$ are periodic with periodicity $2 \pi$, so too are the gauge parameters $\Lambda_0^a$, $\Lambda_0^b$:\footnote{For $p=1$, the gauge parameters $\Lambda_1^a$, $\Lambda_1^b$ may have transition functions between neighboring patches, and these transition functions are periodic. A similar story holds for $p > 1$.}
\begin{equation}
\Lambda_0^a \sim  \Lambda_0^a + 2 \pi \,, ~\Lambda_0^b \sim  \Lambda_0^b + 2 \pi  \,.
\label{periodicg}
\end{equation}
From \eqref{lambdaprime} and \eqref{periodicg}, we see that the periodicities of $\Lambda_0^{a \, \prime}$, $ \Lambda_0^{b \, \prime}$ are more complicated:
\begin{align}
( \Lambda_0^{a \, \prime} ,  \Lambda_0^{b \, \prime} ) \sim (\Lambda_0^{a \, \prime} + 2 \pi \frac{(g_A')^2}{ g_A^2} , \Lambda_0^{b \, \prime}) \sim  (\Lambda_0^{a \, \prime} + 2 \pi \vartheta \frac{(g_A')^2}{g_B^2}, \Lambda_0^{b \, \prime} + 2 \pi )  \,.
\label{periodicities}
\end{align}

Likewise, this kinetic mixing leads to a mixing between the Weak Gravity Conjectures for the gauge fields $A'_p$, $B'_p$. The superextremality condition for a $(p-1)$-brane of charge $(q_A, q_B)$ and tension $T_p$ is given by
\begin{equation}
g_A^2 q_A^2 + g_B^2 q_B^2 \geq T_p^2  M_{\textrm{Pl}; d}^{2-d}\,.
\end{equation}
In the unprimed basis, $q_A$ and $q_B$ are independent integers. However, in the primed basis, the charge lattice is no longer rectangular, and the charges $q_A'$ and $q_B'$ are no longer independent. This is related to the fact that the periodicity conditions \eqref{periodicities} involve a mixing between $\Lambda_p^{a \, \prime}$ and $ \Lambda_p^{b \, \prime}$. The superextremality condition in the primed basis is then given by 
\begin{equation}
\tau^{ij} q_i' q_j' =   (g_B^2 + g_A^2 \vartheta^2) (q_A')^2 + g_A^2 (q_B')^2 + 2 \vartheta g_A^2 q_A' q_B' \geq T_p^2  M_{\textrm{Pl}; d}^{2-d}\,.
\end{equation}
In particular, note the mixing between the two gauge charges via the cross term $2 \vartheta g_A^2 q_A'q_B'$.

Of course, in the case at hand, we may ultimately undo the mixing in the primed basis by rotating back to the unprimed basis. The WGC in theories with multiple gauge fields is a basis-independent statement (sometimes known as the ``convex hull condition'' \cite{Cheung:2014vva}), which holds that every rational direction in the charge lattice must have a superextremal state.\footnote{Here, the tension of a multi-brane state is defined to be the sum of the tensions of the individual branes, $T = \sum_a T_p^a$, and a superextremal multi-brane state is one whose charge-to-tension ratio is greater than that of an extremal black brane of the same charge.} This condition makes decay constant alignment difficult to achieve \cite{rudelius:2015xta, Montero:2015ofa, Brown:2015iha, Heidenreich:2015wga, Heidenreich:2019bjd}.

\section{Conclusions}\label{CONC}

In this paper, we have studied the interplay between emergent higher-group global symmetries, Chern-Simons terms, and Weak Gravity Conjecture mixing. In doing so, we have unified earlier results in axion electrodynamics and extended them to higher dimensions, where supergravity constraints agree perfectly with our expectations. We have learned that the gauge couplings of charged particles and charged strings in higher-dimensional supergravities are related to one another in the presence of certain Chern-Simons couplings, which points us to previously uncovered universal structures within supergravity. Our results offer yet another illustration of the power of generalized global symmetries for constraining and characterizing quantum field theories, including effective field theories in the quantum gravity landscape. 

A number of open questions remain. For one thing, our discussion of axion monodromy ended with a bit of a cliffhanger: it seems that additional UV input is needed if we are to uncover any meaningful constraints among the structures, and it would be worthwhile to explore such UV constraints further. 

Our discussion of kinetic mixing also suggests an interesting possibility: in theories with multiple 1-form gauge fields, the statement of the WGC amounts to the aforementioned ``convex hull condition,'' which depends on the kinetic mixing between various gauge fields. Perhaps, then, there is some analog of the convex hull condition for mixing between gauge fields of different degrees, which depends on the couplings between them? In this work, we have shown that such couplings often mean that the WGC for one gauge field implies the WGC for the other, but we have not entertained the possibility that quantum gravity might impose stronger constraints, similar to the convex hull condition. This possibility is worth exploring further by examining theories in the string landscape.

We have encountered several examples of $C_{d-4} \wedge F_2^+ \wedge F_2^-$ Chern-Simons terms (i.e., terms involving three different gauge fields) for which our usual WGC mixing argument does not apply. Nonetheless, in all of these examples, the string scale for a WGC-satisfying string charged magnetically under $C_{d-4}$ is identified with the scale $\sqrt{g_+ g_-} M_{\text{Pl}; d}^{(d-2)/2}$, so charged string excitations generically satisfy the WGC for either $A^+$ or $A^-$, but not both. It would be interesting to understand this relation and these mixed Chern-Simons couplings in more detail, especially since they show up not only in higher-dimensional supergravity, but also in Kaluza-Klein reductions to four dimensions \cite{Heidenreich:2021yda}, so they may have important consequences for phenomenology. It would also be worthwhile to explore WGC mixing and higher-group symmetries in the presence of even more general Chern-Simons couplings, such as four-term Chern-Simons couplings.

It is encouraging that the Weak Gravity Conjecture has pointed us to universal features of supergravities across diverse dimensions. This suggests that despite the enormous number of studies of the WGC in recent years, its power has not been fully tapped. And, despite decades of study of supergravity, some supergravity stones remain unturned. This gives us hope that other interesting and universal features of supergravity may soon be discovered, and we optimistically anticipate that some of these features will hold up in the quantum gravity landscape even when the assumption of supersymmetry is dropped.

\section*{Acknowledgements}

It is a pleasure to thank Clay C\'ordova, Ben Heidenreich, Jacob McNamara, Kantaro Ohmori, Matthew Reece, Nathan Seiberg, and Ergin Sezgin for useful discussions. We are especially indebted to Matthew Reece for comments on a draft of this manuscript. This work was supported in part by the Berkeley Center for Theoretical Physics; by the Department of Energy, Office of Science, Office of High Energy Physics under QuantISED Award DE-SC0019380 and under contract DE-AC02-05CH11231; and by the National Science Foundation under Award Number 2112880.

\bibliographystyle{utphys}
\bibliography{ref}

\providecommand{\href}[2]{#2}\begingroup\raggedright\begin{thebibliography}{10}

\bibitem{Gaiotto:2014kfa}
D.~Gaiotto, A.~Kapustin, N.~Seiberg, and B.~Willett, ``{Generalized Global
  Symmetries},'' \href{http://dx.doi.org/10.1007/JHEP02(2015)172}{{\em JHEP}
  {\bfseries 02} (2015) 172},
\href{http://arxiv.org/abs/1412.5148}{{\ttfamily arXiv:1412.5148 [hep-th]}}.

\bibitem{Seiberg:2019vrp}
N.~Seiberg, ``{Field Theories With a Vector Global Symmetry},''
  \href{http://dx.doi.org/10.21468/SciPostPhys.8.4.050}{{\em SciPost Phys.}
  {\bfseries 8} (2020) 050}, \href{http://arxiv.org/abs/1909.10544}{{\ttfamily
  arXiv:1909.10544 [cond-mat.str-el]}}.

\bibitem{McNamara:2019rup}
J.~McNamara and C.~Vafa, ``{Cobordism Classes and the Swampland},''
  \href{http://arxiv.org/abs/1909.10355}{{\ttfamily arXiv:1909.10355
  [hep-th]}}.

\bibitem{Sharpe:2015mja}
E.~Sharpe, ``{Notes on generalized global symmetries in QFT},''
  \href{http://dx.doi.org/10.1002/prop.201500048}{{\em Fortsch. Phys.}
  {\bfseries 63} (2015) 659--682},
  \href{http://arxiv.org/abs/1508.04770}{{\ttfamily arXiv:1508.04770
  [hep-th]}}.

\bibitem{Cordova:2018cvg}
C.~C\'ordova, T.~T. Dumitrescu, and K.~Intriligator, ``{Exploring 2-Group
  Global Symmetries},'' \href{http://dx.doi.org/10.1007/JHEP02(2019)184}{{\em
  JHEP} {\bfseries 02} (2019) 184},
  \href{http://arxiv.org/abs/1802.04790}{{\ttfamily arXiv:1802.04790
  [hep-th]}}.

\bibitem{Benini:2018reh}
F.~Benini, C.~C\'ordova, and P.-S. Hsin, ``{On 2-Group Global Symmetries and
  their Anomalies},'' \href{http://dx.doi.org/10.1007/JHEP03(2019)118}{{\em
  JHEP} {\bfseries 03} (2019) 118},
  \href{http://arxiv.org/abs/1803.09336}{{\ttfamily arXiv:1803.09336
  [hep-th]}}.

\bibitem{Hawking:1974sw}
S.~Hawking, ``{Particle Creation by Black Holes},''
  \href{http://dx.doi.org/10.1007/BF02345020}{{\em Commun. Math. Phys.}
  {\bfseries 43} (1975) 199--220}. [Erratum: Commun.Math.Phys. 46, 206 (1976)].

\bibitem{Zeldovich:1976vq}
Y.~B. Zeldovich, ``{A New Type of Radioactive Decay: Gravitational Annihilation
  of Baryons},'' \href{http://dx.doi.org/10.1016/0375-9601(76)90783-0}{{\em
  Phys. Lett. A} {\bfseries 59} (1976) 254}.

\bibitem{Zeldovich:1977be}
Y.~B. Zeldovich, ``{A Novel Type of Radioactive Decay: Gravitational Baryon
  Annihilation},'' {\em Zh. Eksp. Teor. Fiz.} {\bfseries 72} (1977) 18--21.

\bibitem{Banks:1988yz}
T.~Banks and L.~J. Dixon, ``{Constraints on String Vacua with Space-Time
  Supersymmetry},''
\href{http://dx.doi.org/10.1016/0550-3213(88)90523-8}{{\em Nucl. Phys.}
  {\bfseries B307} (1988) 93--108}.

\bibitem{Giddings:1988cx}
S.~B. Giddings and A.~Strominger, ``{Loss of Incoherence and Determination of
  Coupling Constants in Quantum Gravity},''
  \href{http://dx.doi.org/10.1016/0550-3213(88)90109-5}{{\em Nucl. Phys. B}
  {\bfseries 307} (1988) 854--866}.

\bibitem{Abbott:1989jw}
L.~F. Abbott and M.~B. Wise, ``{Wormholes and Global Symmetries},''
  \href{http://dx.doi.org/10.1016/0550-3213(89)90503-8}{{\em Nucl. Phys. B}
  {\bfseries 325} (1989) 687--704}.

\bibitem{Coleman:1989zu}
S.~R. Coleman and K.-M. Lee, ``{Wormholes Made Without Massless Matter
  Fields},'' \href{http://dx.doi.org/10.1016/0550-3213(90)90149-8}{{\em Nucl.
  Phys. B} {\bfseries 329} (1990) 387--409}.

\bibitem{Kallosh:1995hi}
R.~Kallosh, A.~D. Linde, D.~A. Linde, and L.~Susskind, ``{Gravity and global
  symmetries},'' \href{http://dx.doi.org/10.1103/PhysRevD.52.912}{{\em
  Phys.Rev.} {\bfseries D52} (1995) 912--935},
\href{http://arxiv.org/abs/hep-th/9502069}{{\ttfamily arXiv:hep-th/9502069
  [hep-th]}}.

\bibitem{Witten:1998qj}
E.~Witten, ``{Anti-de Sitter space and holography},''
  \href{http://dx.doi.org/10.4310/ATMP.1998.v2.n2.a2}{{\em Adv. Theor. Math.
  Phys.} {\bfseries 2} (1998) 253--291},
  \href{http://arxiv.org/abs/hep-th/9802150}{{\ttfamily arXiv:hep-th/9802150}}.

\bibitem{Banks:2010zn}
T.~Banks and N.~Seiberg, ``{Symmetries and Strings in Field Theory and
  Gravity},'' \href{http://dx.doi.org/10.1103/PhysRevD.83.084019}{{\em Phys.
  Rev.} {\bfseries D83} (2011) 084019},
\href{http://arxiv.org/abs/1011.5120}{{\ttfamily arXiv:1011.5120 [hep-th]}}.

\bibitem{Harlow:2018tng}
D.~Harlow and H.~Ooguri, ``{Symmetries in quantum field theory and quantum
  gravity},'' \href{http://dx.doi.org/10.1007/s00220-021-04040-y}{{\em Commun.
  Math. Phys.} {\bfseries 383} no.~3, (2021) 1669--1804},
  \href{http://arxiv.org/abs/1810.05338}{{\ttfamily arXiv:1810.05338
  [hep-th]}}.

\bibitem{Harlow:2020bee}
D.~Harlow and E.~Shaghoulian, ``{Global symmetry, Euclidean gravity, and the
  black hole information problem},''
  \href{http://dx.doi.org/10.1007/JHEP04(2021)175}{{\em JHEP} {\bfseries 04}
  (2021) 175}, \href{http://arxiv.org/abs/2010.10539}{{\ttfamily
  arXiv:2010.10539 [hep-th]}}.

\bibitem{Chen:2020ojn}
Y.~Chen and H.~W. Lin, ``{Signatures of global symmetry violation in relative
  entropies and replica wormholes},''
  \href{http://dx.doi.org/10.1007/JHEP03(2021)040}{{\em JHEP} {\bfseries 03}
  (2021) 040}, \href{http://arxiv.org/abs/2011.06005}{{\ttfamily
  arXiv:2011.06005 [hep-th]}}.

\bibitem{Hsin:2020mfa}
P.-S. Hsin, L.~V. Iliesiu, and Z.~Yang, ``{A violation of global symmetries
  from replica wormholes and the fate of black hole remnants},''
  \href{http://dx.doi.org/10.1088/1361-6382/ac2134}{{\em Class. Quant. Grav.}
  {\bfseries 38} no.~19, (2021) 194004},
  \href{http://arxiv.org/abs/2011.09444}{{\ttfamily arXiv:2011.09444
  [hep-th]}}.

\bibitem{Yonekura:2020ino}
K.~Yonekura, ``{Topological violation of global symmetries in quantum
  gravity},'' \href{http://arxiv.org/abs/2011.11868}{{\ttfamily
  arXiv:2011.11868 [hep-th]}}.

\bibitem{Rudelius:2020orz}
T.~Rudelius and S.-H. Shao, ``{Topological Operators and Completeness of
  Spectrum in Discrete Gauge Theories},''
  \href{http://dx.doi.org/10.1007/JHEP12(2020)172}{{\em JHEP} {\bfseries 12}
  (2020) 172}, \href{http://arxiv.org/abs/2006.10052}{{\ttfamily
  arXiv:2006.10052 [hep-th]}}.

\bibitem{Heidenreich:2020pkc}
B.~Heidenreich, J.~McNamara, M.~Montero, M.~Reece, T.~Rudelius, and
  I.~Valenzuela, ``{Chern-Weil Global Symmetries and How Quantum Gravity Avoids
  Them},'' \href{http://arxiv.org/abs/2012.00009}{{\ttfamily arXiv:2012.00009
  [hep-th]}}.

\bibitem{polchinski:2003bq}
J.~Polchinski, ``{Monopoles, duality, and string theory},''
  \href{http://dx.doi.org/10.1142/S0217751X0401866X}{{\em Int.J.Mod.Phys.}
  {\bfseries A19S1} (2004) 145--156},
\href{http://arxiv.org/abs/hep-th/0304042}{{\ttfamily arXiv:hep-th/0304042
  [hep-th]}}.

\bibitem{Heidenreich:2021tna}
B.~Heidenreich, J.~Mcnamara, M.~Montero, M.~Reece, T.~Rudelius, and
  I.~Valenzuela, ``{Non-Invertible Global Symmetries and Completeness of the
  Spectrum},'' \href{http://arxiv.org/abs/2104.07036}{{\ttfamily
  arXiv:2104.07036 [hep-th]}}.

\bibitem{Casini:2021zgr}
H.~Casini and J.~M. Magan, ``{On completeness and generalized symmetries in
  quantum field theory},''
  \href{http://dx.doi.org/10.1142/S0217732321300251}{{\em Mod. Phys. Lett. A}
  {\bfseries 36} no.~36, (2021) 2130025},
  \href{http://arxiv.org/abs/2110.11358}{{\ttfamily arXiv:2110.11358
  [hep-th]}}.

\bibitem{Nomura:2019qps}
Y.~Nomura, ``{Spacetime and Universal Soft Modes --- Black Holes and Beyond},''
  \href{http://dx.doi.org/10.1103/PhysRevD.101.066024}{{\em Phys. Rev. D}
  {\bfseries 101} no.~6, (2020) 066024},
  \href{http://arxiv.org/abs/1908.05728}{{\ttfamily arXiv:1908.05728
  [hep-th]}}.

\bibitem{Fichet:2019ugl}
S.~Fichet and P.~Saraswat, ``{Approximate Symmetries and Gravity},''
  \href{http://dx.doi.org/10.1007/JHEP01(2020)088}{{\em JHEP} {\bfseries 01}
  (2020) 088}, \href{http://arxiv.org/abs/1909.02002}{{\ttfamily
  arXiv:1909.02002 [hep-th]}}.

\bibitem{CordovaOhmoriRudelius}
C.~C\'ordova, K.~Ohmori, and T.~Rudelius, ``{Symmetry Breaking Scales and Weak
  Gravity Conjectures},'' 2022.
\newblock to appear.

\bibitem{Daus:2020vtf}
T.~Daus, A.~Hebecker, S.~Leonhardt, and J.~March-Russell, ``{Towards a
  Swampland Global Symmetry Conjecture using weak gravity},''
  \href{http://dx.doi.org/10.1016/j.nuclphysb.2020.115167}{{\em Nucl. Phys. B}
  {\bfseries 960} (2020) 115167},
  \href{http://arxiv.org/abs/2002.02456}{{\ttfamily arXiv:2002.02456
  [hep-th]}}.

\bibitem{Arkanihamed:2006dz}
N.~Arkani-Hamed, L.~Motl, A.~Nicolis, and C.~Vafa, ``{The String landscape,
  black holes and gravity as the weakest force},''
  \href{http://dx.doi.org/10.1088/1126-6708/2007/06/060}{{\em JHEP} {\bfseries
  0706} (2007) 060},
\href{http://arxiv.org/abs/hep-th/0601001}{{\ttfamily arXiv:hep-th/0601001
  [hep-th]}}.

\bibitem{Harlow:2022gzl}
D.~Harlow, B.~Heidenreich, M.~Reece, and T.~Rudelius, ``{The Weak Gravity
  Conjecture: A Review},'' \href{http://arxiv.org/abs/2201.08380}{{\ttfamily
  arXiv:2201.08380 [hep-th]}}.

\bibitem{Brennan:2020ehu}
T.~D. Brennan and C.~Cordova, ``{Axions, Higher-Groups, and Emergent
  Symmetry},'' \href{http://arxiv.org/abs/2011.09600}{{\ttfamily
  arXiv:2011.09600 [hep-th]}}.

\bibitem{Heidenreich:2021yda}
B.~Heidenreich, M.~Reece, and T.~Rudelius, ``{The Weak Gravity Conjecture and
  Axion Strings},'' \href{http://arxiv.org/abs/2108.11383}{{\ttfamily
  arXiv:2108.11383 [hep-th]}}.

\bibitem{McAllister:2008hb}
L.~McAllister, E.~Silverstein, and A.~Westphal, ``{Gravity Waves and Linear
  Inflation from Axion Monodromy},''
  \href{http://dx.doi.org/10.1103/PhysRevD.82.046003}{{\em Phys. Rev. D}
  {\bfseries 82} (2010) 046003},
  \href{http://arxiv.org/abs/0808.0706}{{\ttfamily arXiv:0808.0706 [hep-th]}}.

\bibitem{Silverstein:2008sg}
E.~Silverstein and A.~Westphal, ``{Monodromy in the CMB: Gravity Waves and
  String Inflation},'' \href{http://dx.doi.org/10.1103/PhysRevD.78.106003}{{\em
  Phys. Rev. D} {\bfseries 78} (2008) 106003},
  \href{http://arxiv.org/abs/0803.3085}{{\ttfamily arXiv:0803.3085 [hep-th]}}.

\bibitem{kim:2004rp}
J.~E. Kim, H.~P. Nilles, and M.~Peloso, ``{Completing natural inflation},''
  \href{http://dx.doi.org/10.1088/1475-7516/2005/01/005}{{\em JCAP} {\bfseries
  0501} (2005) 005},
\href{http://arxiv.org/abs/hep-ph/0409138}{{\ttfamily arXiv:hep-ph/0409138
  [hep-ph]}}.

\bibitem{Bhardwaj:2017xup}
L.~Bhardwaj and Y.~Tachikawa, ``{On finite symmetries and their gauging in two
  dimensions},'' \href{http://dx.doi.org/10.1007/JHEP03(2018)189}{{\em JHEP}
  {\bfseries 03} (2018) 189}, \href{http://arxiv.org/abs/1704.02330}{{\ttfamily
  arXiv:1704.02330 [hep-th]}}.

\bibitem{Chang:2018iay}
C.-M. Chang, Y.-H. Lin, S.-H. Shao, Y.~Wang, and X.~Yin, ``{Topological Defect
  Lines and Renormalization Group Flows in Two Dimensions},''
  \href{http://dx.doi.org/10.1007/JHEP01(2019)026}{{\em JHEP} {\bfseries 01}
  (2019) 026}, \href{http://arxiv.org/abs/1802.04445}{{\ttfamily
  arXiv:1802.04445 [hep-th]}}.

\bibitem{Thorngren:2019iar}
R.~Thorngren and Y.~Wang, ``{Fusion Category Symmetry I: Anomaly In-Flow and
  Gapped Phases},'' \href{http://arxiv.org/abs/1912.02817}{{\ttfamily
  arXiv:1912.02817 [hep-th]}}.

\bibitem{Sharpe:2021srf}
E.~Sharpe, ``{Topological operators, noninvertible symmetries and
  decomposition},'' \href{http://arxiv.org/abs/2108.13423}{{\ttfamily
  arXiv:2108.13423 [hep-th]}}.

\bibitem{Koide:2021zxj}
M.~Koide, Y.~Nagoya, and S.~Yamaguchi, ``{Non-invertible topological defects in
  4-dimensional $\mathbb{Z}_2$ pure lattice gauge theory},''
  \href{http://arxiv.org/abs/2109.05992}{{\ttfamily arXiv:2109.05992
  [hep-th]}}.

\bibitem{Choi:2021kmx}
Y.~Choi, C.~Cordova, P.-S. Hsin, H.~T. Lam, and S.-H. Shao, ``{Non-Invertible
  Duality Defects in 3+1 Dimensions},''
  \href{http://arxiv.org/abs/2111.01139}{{\ttfamily arXiv:2111.01139
  [hep-th]}}.

\bibitem{Tanizaki:2019rbk}
Y.~Tanizaki and M.~{\"U}nsal, ``{Modified instanton sum in QCD and
  higher-groups},'' \href{http://dx.doi.org/10.1007/JHEP03(2020)123}{{\em JHEP}
  {\bfseries 03} (2020) 123}, \href{http://arxiv.org/abs/1912.01033}{{\ttfamily
  arXiv:1912.01033 [hep-th]}}.

\bibitem{Fan:2021ntg}
J.~Fan, K.~Fraser, M.~Reece, and J.~Stout, ``{Axion Mass from Magnetic Monopole
  Loops},'' \href{http://arxiv.org/abs/2105.09950}{{\ttfamily arXiv:2105.09950
  [hep-ph]}}.

\bibitem{Reece:2018zvv}
M.~Reece, ``{Photon Masses in the Landscape and the Swampland},''
  \href{http://dx.doi.org/10.1007/JHEP07(2019)181}{{\em JHEP} {\bfseries 07}
  (2019) 181}, \href{http://arxiv.org/abs/1808.09966}{{\ttfamily
  arXiv:1808.09966 [hep-th]}}.

\bibitem{Heidenreich:2015nta}
B.~Heidenreich, M.~Reece, and T.~Rudelius, ``{Sharpening the Weak Gravity
  Conjecture with Dimensional Reduction},''
  \href{http://dx.doi.org/10.1007/JHEP02(2016)140}{{\em JHEP} {\bfseries 02}
  (2016) 140},
\href{http://arxiv.org/abs/1509.06374}{{\ttfamily arXiv:1509.06374 [hep-th]}}.

\bibitem{Heidenreich:2016aqi}
B.~Heidenreich, M.~Reece, and T.~Rudelius, ``{Evidence for a Lattice Weak
  Gravity Conjecture},'' \href{http://dx.doi.org/10.1007/JHEP08(2017)025}{{\em
  JHEP} {\bfseries 08} (2017) 025},
\href{http://arxiv.org/abs/1606.08437}{{\ttfamily arXiv:1606.08437 [hep-th]}}.

\bibitem{Andriolo:2018lvp}
S.~Andriolo, D.~Junghans, T.~Noumi, and G.~Shiu, ``{A Tower Weak Gravity
  Conjecture from Infrared Consistency},''
  \href{http://dx.doi.org/10.1002/prop.201800020}{{\em Fortsch. Phys.}
  {\bfseries 66} no.~5, (2018) 1800020},
\href{http://arxiv.org/abs/1802.04287}{{\ttfamily arXiv:1802.04287 [hep-th]}}.

\bibitem{Kats:2006xp}
Y.~Kats, L.~Motl, and M.~Padi, ``{Higher-order corrections to mass-charge
  relation of extremal black holes},''
  \href{http://dx.doi.org/10.1088/1126-6708/2007/12/068}{{\em JHEP} {\bfseries
  12} (2007) 068}, \href{http://arxiv.org/abs/hep-th/0606100}{{\ttfamily
  arXiv:hep-th/0606100}}.

\bibitem{Callan:1984sa}
C.~G. Callan, Jr. and J.~A. Harvey, ``{Anomalies and Fermion Zero Modes on
  Strings and Domain Walls},''
  \href{http://dx.doi.org/10.1016/0550-3213(85)90489-4}{{\em Nucl. Phys. B}
  {\bfseries 250} (1985) 427--436}.

\bibitem{Heidenreich:2017sim}
B.~Heidenreich, M.~Reece, and T.~Rudelius, ``{The Weak Gravity Conjecture and
  Emergence from an Ultraviolet Cutoff},''
  \href{http://dx.doi.org/10.1140/epjc/s10052-018-5811-3}{{\em Eur. Phys. J.}
  {\bfseries C78} no.~4, (2018) 337},
\href{http://arxiv.org/abs/1712.01868}{{\ttfamily arXiv:1712.01868 [hep-th]}}.

\bibitem{BPSstrings}
B.~Heidenreich and T.~Rudelius, ``The {Weak Gravity Conjecture and BPS
  Strings},'' 2022.
\newblock to appear.

\bibitem{Riccioni:1999xq}
F.~Riccioni, ``{Abelian vector multiplets in six-dimensional supergravity},''
  \href{http://dx.doi.org/10.1016/S0370-2693(00)00003-4}{{\em Phys. Lett. B}
  {\bfseries 474} (2000) 79--84},
  \href{http://arxiv.org/abs/hep-th/9910246}{{\ttfamily arXiv:hep-th/9910246}}.

\bibitem{Sagnotti:1992qw}
A.~Sagnotti, ``{A Note on the Green-Schwarz mechanism in open string
  theories},'' \href{http://dx.doi.org/10.1016/0370-2693(92)90682-T}{{\em Phys.
  Lett. B} {\bfseries 294} (1992) 196--203},
  \href{http://arxiv.org/abs/hep-th/9210127}{{\ttfamily arXiv:hep-th/9210127}}.

\bibitem{Riccioni:2001bg}
F.~Riccioni, ``{All couplings of minimal six-dimensional supergravity},''
  \href{http://dx.doi.org/10.1016/S0550-3213(01)00199-7}{{\em Nucl. Phys. B}
  {\bfseries 605} (2001) 245--265},
  \href{http://arxiv.org/abs/hep-th/0101074}{{\ttfamily arXiv:hep-th/0101074}}.

\bibitem{Lee:2019xtm}
S.-J. Lee, W.~Lerche, and T.~Weigand, ``{Emergent Strings, Duality and Weak
  Coupling Limits for Two-Form Fields},''
  \href{http://arxiv.org/abs/1904.06344}{{\ttfamily arXiv:1904.06344
  [hep-th]}}.

\bibitem{Bergshoeff:1985mr}
E.~Bergshoeff, I.~G. Koh, and E.~Sezgin, ``{{Yang-Mills} / Einstein
  Supergravity in Seven-dimensions},''
  \href{http://dx.doi.org/10.1103/PhysRevD.32.1353}{{\em Phys. Rev. D}
  {\bfseries 32} (1985) 1353--1357}.

\bibitem{Awada:1985ag}
M.~Awada and P.~K. Townsend, ``{d = 8 MAXWELL-EINSTEIN SUPERGRAVITY},''
  \href{http://dx.doi.org/10.1016/0370-2693(85)91353-X}{{\em Phys. Lett. B}
  {\bfseries 156} (1985) 51--54}.

\bibitem{Polchinski:1998rr}
J.~Polchinski, \href{http://dx.doi.org/10.1017/CBO9780511618123}{{\em {String
  theory. Vol. 2: Superstring theory and beyond}}}.
\newblock Cambridge Monographs on Mathematical Physics. Cambridge University
  Press, 12, 2007.

\bibitem{Dolan:2017vmn}
M.~J. Dolan, P.~Draper, J.~Kozaczuk, and H.~Patel, ``{Transplanckian Censorship
  and Global Cosmic Strings},''
  \href{http://dx.doi.org/10.1007/JHEP04(2017)133}{{\em JHEP} {\bfseries 04}
  (2017) 133},
\href{http://arxiv.org/abs/1701.05572}{{\ttfamily arXiv:1701.05572 [hep-th]}}.

\bibitem{Hebecker:2017wsu}
A.~Hebecker, P.~Henkenjohann, and L.~T. Witkowski, ``{What is the Magnetic Weak
  Gravity Conjecture for Axions?},''
  \href{http://dx.doi.org/10.1002/prop.201700011}{{\em Fortsch. Phys.}
  {\bfseries 65} no.~3-4, (2017) 1700011},
\href{http://arxiv.org/abs/1701.06553}{{\ttfamily arXiv:1701.06553 [hep-th]}}.

\bibitem{Kaloper:2008fb}
N.~Kaloper and L.~Sorbo, ``{A Natural Framework for Chaotic Inflation},''
  \href{http://dx.doi.org/10.1103/PhysRevLett.102.121301}{{\em Phys. Rev.
  Lett.} {\bfseries 102} (2009) 121301},
  \href{http://arxiv.org/abs/0811.1989}{{\ttfamily arXiv:0811.1989 [hep-th]}}.

\bibitem{Cheung:2014vva}
C.~Cheung and G.~N. Remmen, ``{Naturalness and the Weak Gravity Conjecture},''
  \href{http://dx.doi.org/10.1103/PhysRevLett.113.051601}{{\em Phys.Rev.Lett.}
  {\bfseries 113} (2014) 051601},
\href{http://arxiv.org/abs/1402.2287}{{\ttfamily arXiv:1402.2287 [hep-ph]}}.

\bibitem{rudelius:2015xta}
T.~Rudelius, ``{Constraints on Axion Inflation from the Weak Gravity
  Conjecture},'' \href{http://dx.doi.org/10.1088/1475-7516/2015/9/020}{{\em
  JCAP} {\bfseries 09} (2015) 020},
\href{http://arxiv.org/abs/1503.00795}{{\ttfamily arXiv:1503.00795 [hep-th]}}.

\bibitem{Montero:2015ofa}
M.~Montero, A.~M. Uranga, and I.~Valenzuela, ``{Transplanckian axions!?},''
  \href{http://dx.doi.org/10.1007/JHEP08(2015)032}{{\em JHEP} {\bfseries 08}
  (2015) 032}, \href{http://arxiv.org/abs/1503.03886}{{\ttfamily
  arXiv:1503.03886 [hep-th]}}.

\bibitem{Brown:2015iha}
J.~Brown, W.~Cottrell, G.~Shiu, and P.~Soler, ``{Fencing in the Swampland:
  Quantum Gravity Constraints on Large Field Inflation},''
  \href{http://dx.doi.org/10.1007/JHEP10(2015)023}{{\em JHEP} {\bfseries 10}
  (2015) 023}, \href{http://arxiv.org/abs/1503.04783}{{\ttfamily
  arXiv:1503.04783 [hep-th]}}.

\bibitem{Heidenreich:2015wga}
B.~Heidenreich, M.~Reece, and T.~Rudelius, ``{Weak Gravity Strongly Constrains
  Large-Field Axion Inflation},''
  \href{http://dx.doi.org/10.1007/JHEP12(2015)108}{{\em JHEP} {\bfseries 12}
  (2015) 108},
\href{http://arxiv.org/abs/1506.03447}{{\ttfamily arXiv:1506.03447 [hep-th]}}.

\bibitem{Heidenreich:2019bjd}
B.~Heidenreich, C.~Long, L.~McAllister, T.~Rudelius, and J.~Stout, ``{Instanton
  Resummation and the Weak Gravity Conjecture},''
  \href{http://dx.doi.org/10.1007/JHEP11(2020)166}{{\em JHEP} {\bfseries 11}
  (2020) 166}, \href{http://arxiv.org/abs/1910.14053}{{\ttfamily
  arXiv:1910.14053 [hep-th]}}.

\end{thebibliography}\endgroup
\end{document}